\documentclass[letterpaper, 10 pt, conference]{ieeeconf}  

\IEEEoverridecommandlockouts                              

\usepackage{graphics} 
\usepackage{epsfig} 
\usepackage{mathptmx} 
\usepackage{times} 
\usepackage{amsmath} 
\usepackage{amssymb}  
\usepackage{float}
\newtheorem{thm}{Theorem}
\newtheorem{remark}{Remark}
\newtheorem{claim}{Claim}
\usepackage{color}
\usepackage{subcaption}
\usepackage{hyperref}
\usepackage{tcolorbox}

\title{\LARGE \bf
Control Synthesis for an Underactuated Cable Suspended System Using Dynamic Decoupling }

\author{Siddharth H. Nair$^{1}$, Ravi N. Banavar$^{2}$ and D.H.S Maithripala$^{3}$
\thanks{$^{1}$Siddharth H. Nair is with the Department of Aerospace Engineering, Indian Institute of Technology Bombay, Mumbai 400076, India
{\tt\small siddharth.nair@iitb.ac.in}}%
\thanks{$^{2}$Ravi N. Banavar is with the Faculty of Systems and Control Engineering, Indian Institute of Technology Bombay, Mumbai 400076, India
{\tt\small banavar@iitb.ac.in}}%
\thanks{$^{2}$D.H.S Maithripala is with the Faculty of Mechanical Engineering, University of Peradeniya, KY 20400, Sri Lanka
{\tt\small smaithri@pdn.ac.lk}}%
}

\begin{document}

\maketitle
\thispagestyle{empty}
\pagestyle{empty}

\begin{abstract}
This article studies the dynamics and control of a novel underactuated system, wherein a plate suspended
by cables and with a freely moving mass on top, whose other ends are attached to three quadrotors, is 
sought to be horizontally stabilized at a certain height, with the ball positioned at the center of mass of the plate.  
The freely moving mass introduces a 2-degree of underactuation into the system. The design proceeds through a decoupling
of the quadrotors and the plate dynamics. Through a partial feedback linearization approach, the attitude of the plate and
 the translational height of the plate is initially controlled, while maintaining a bounded velocity along the $y$ and $x$
 directions. These inputs are then synthesized through the quadrotors with a backstepping and 
 timescale separation argument based on Tikhonov's theorem.
\end{abstract}
\section{Introduction}
Quadrotor drones are increasingly gaining popularity in non-military security
applications like surveillance, communication relays and civil applications
like environmental monitoring, traffic control, disaster relief and construction
\cite{beard}.
Trajectory tracking controllers for quadrotors have been successfully studied in \cite{leetrack,hofftrack,gtrack,maithripalatrack} while control of quadrotor formations has been studied in \cite{vkform,vkform2,maithripalaform}.
Recent endeavours involve synthesizing control laws wherein quadrotors are required
to transport loads from one point to another. In \cite{goodarzi}, a system consisting of a quadrotor
and a flexible cable treated as serially-connected links is modelled in a coordinate-free
form where the equations of motion are obtained using infinitesimal variations
of elements belonging to a Lie group. First the desired forces on the links are derived so that the payload tracks a desired trajectory and next the thrust and moments acting on the quadrotor are derived so that these forces are in turn, generated by the quadrotor. In \cite{leevk}, multiple quadrotors
carrying a point payload via rigid, massless links is considered while in \cite{lee}, the suspended payload is a rigid body. The design philosophy is similar to that used in \cite{goodarzi}- where they first design the desired forces in the links and then use the quadrotors to generate them. This work is extended in \cite{goodarzitl} to incorporate
flexible cables. 
The development of such systems can find applications in real life situations like transportation and search and rescue operations.\\\par
Much of the previous effort in the field of cable suspended systems focusses on fully actuated 
systems. Here, we introduce an element of underactuation into the system in the form of a freely moving ball on a
plate, thereby increasing the control complexity of the problem. Designing control strategies to accommodate for underactuated payloads facilitates design of simpler mechanisms for transporting payload of various geometries. For example, a simple, flat platform would require a payload to comply with fewer geometric constraints than say, a sophisticated mechanical gripper. Secondly, the underactuated nature of the payload system adds inherent consideration for transporting delicate payloads (a fluid container, for instance). The work \cite{rubio} considers 1-D version of the problem of balancing a ball on a rod connected to two quadrotors via rigid
links. The quadrotors are restricted to move only vertically and the stabilization of the ball is achieved by employing a model predictive controller for the linearized model of the system. In this article, we consider the full 3-D problem.
The control approach we adopt can be summarized as follows: First, the quadrotors are decoupled from the ball-and-plate system and then the desired forces in the tethers are synthesized such that the control objectives are met. The forces in the tethers create both the force and the
torque to position and orient the plate. These forces are then generated by the respective quadrotors using a backstepping-like strategy seen in \cite{lee}. The response of the quadrotors are assumed significantly faster than the ball-and-plate dynamics. 
The underactuated subsystem leads us to employ partial feedback linearization into our control design.\\
\par
The remaining article is organized as follows. Section \ref{pf} formally sets up the problem by describing the system of interest, fixing up naming conventions and deriving a coordinate-form of the equations of motion using Lagrangian mechanics. In section \ref{cd}, the control systems are constructed followed by numerical validation via simulations in section \ref{sim}

\section{Problem Formulation}\label{pf}
Consider three quadrotors with masses $m_1,\ m_2$ and $m_3$ and inertias $J_1,\ J_2$ and $J_3$ respectively, carrying a thin plate of mass $m_p$ and inertia $J_p$ via three inextensible cables of lengths $l_1,\ l_2$ and $l_3$ respectively. The plate also carries a ball of mass $m_b$. The inertial coordinate system is set up as shown in figure \ref{f:sys}.
\begin{figure}[H]
\centering
\includegraphics[scale=0.3]{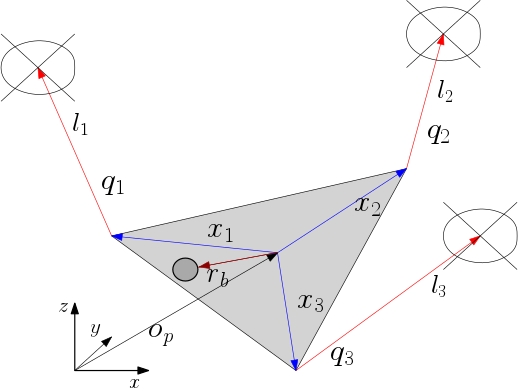}
\caption{The figure depicts a ball on a plate system slung by three quadrotors.}
\label{f:sys}
\end{figure}
The location of the centre of mass of the plate in the inertial frame is denoted by $o_p \in \mathbb{R}^3$ and its attitude is denoted by $R_p \in \mathop{SO(3)}$. Let $x_i \in \mathbb{R}^3$ be the vector from the centre of mass of plate to the point where the $i$-th cable is attached to the plate. These vectors are constant and lie in the coordinate system attached to the plate at its centre of mass. The vector $r_b \in \mathbb{R}^2$ represents the position of the ball on the plate. To represent $r_b$ as a 3 dimensional vector, we define the matrix $E=\begin{bmatrix} 1 & 0\\ 0& 1\\ 0& 0\end{bmatrix}$, which when multiplied by $r_b$ yields the ball's position in the plate's reference frame. In the inertial frame, the position of the ball is given by $o_b=o_p+R_pEr_b$.\\
\par
Since the tethers are inextensible, the locus of the positions of the $i$-th quadrotor is a sphere of radius $l_i$ and centered at $x_i$ in the plate's coordinate system. Thus, its position in the inertial coordinate system is given by $o_i=o_p+R_px_i+l_iq_i$ where $q_i \in \mathop{S}^2$ is a unit vector aligned along the $i$-th cable. Assuming that the tethers aren't hinged to the quadrotors rigidly, the attitude of the $i$-th quadrotor is denoted by $R_i \in \mathop{SO(3)}$. To consolidate, the states of the system as a whole evolve over the configuration manifold $\mathbf{Q}$ given by $$\mathbf{Q}=\underbrace{\mathop{SO(3)}\times\mathbb{R}^3}_{Plate}\times\underbrace{\mathbb{R}^2}_{Ball}\times \underbrace{(\mathop{S^2}\times\mathop{SO(3)})^3}_{Quadrotors}$$The coordinates $(\underbrace{(o_p,R_p)}_{Plate},\underbrace{r_b}_{ball},\underbrace{((q_1,R_1),(q_2,R_2),(q_3,R_3))}_{Quadrotors})$ describe the state of the system.\\
\par
As the name suggests, each quadrotor is propelled using 4 motors which generate a net thrust along the yaw axis and a moment. Let the thrust force and moment generated by the $i$-th quadrotor in its coordinate system be given by $f_ie_3$ and $M_i$ respectively where $f_i$ is the magnitude of the thrust and $e_3$ is the unit vector aligned along the quadrotor's yaw axis. Thus $\{f_i,M_i\}_{i=1,2,3}$ are the control inputs to the system.\\
\par
The objective is to design $\{f_i,M_i\}_{i=1,2,3}$ for the quadrotor so as to stabilize the attitude at $R_p=I$ and the height of the centre of mass of the plate while simultaneously stabilizing the ball at the plate's centre of mass.
\subsection{Equations of Motion}
The rotational kinematics of the $i$th link, quadrotor and plate, respectively, are given by 
\begin{align}
\dot{q}_i&=\omega_i\times q_i=\hat{\omega}_iq_i\\
\dot{R}_i&=R_i\hat{\Omega}_i\\
\dot{R}_p&=R_p\hat{\Omega}_p
\end{align}
where $\omega_i$ is the angular velocity of the tether expressed in the inertial frame. The tether
is assumed to be taut and further, there is no component of its angular velocity along the tether axis. We
express this as $\omega_i \cdot q_i=0$.  $\Omega_p$ and $\Omega_i$ are the angular velocities of the  plate and $i$-th quadrotor respectively in their respective coordinate systems, and the operator\ $\hat{\cdot}$ is the map from $\mathbb{R}^3$ to the space of skew-symmetric matrices as defined by $$\hat{x}=\begin{bmatrix}0& -x_3 & x_2\\x_3&0&-x_1\\-x_2&x_1&0\end{bmatrix}$$ for $x=[x_1\ x_2\ x_3]^T \in \mathbb{R}^3$.
\par
The equations of motion are derived using a variational approach. The kinetic energy and potential energy of the system are given by 
\normalsize{\begin{align*}
\mathcal{T}=&\underbrace{\frac{1}{2}m_p||\dot{o}_p||^2+\frac{1}{2}\Omega^T_pJ_p\Omega_p}_{Plate}\nonumber\\
&+\underbrace{\sum_{i=1}^3\frac{1}{2}m_i||\dot{o}_p+R_p\hat{\Omega}_px_i+l_i\hat{\omega}_iq_i||^2+\frac{1}{2}\Omega^T_iJ_i\Omega_i}_{Quadrotors}\nonumber\\
&+\underbrace{\frac{1}{2}m_b||\dot{o}_p+R_p\hat{\Omega}_pEr_b+R_pE\dot{r}_b||^2}_{Ball}\\
\mathcal{U}=&\underbrace{m_pge^T_3o_p}_{Plate}+\underbrace{\sum_{i=1}^3m_ige^T_3(o_p+R_px_i+l_iq_i)}_{Quadrotors}\nonumber\\
&+\underbrace{m_bge^T_3(o_p+R_pEr_b)}_{Ball}
\end{align*}}

The Lagrangian $\mathcal{L}:\mathbf{Q}\rightarrow\mathbb{R}$ of the system is obtained as the difference between the kinetic and potential energies, i.e, $\mathcal{L}=\mathcal{T}-\mathcal{U}$. The action integral is given by $\mathcal{A}=\int_{t_0}^{t_f}\mathcal{L}dt$. Further, we define the differential operator $\mathbf{D}_{u}(\cdot): \mathbb{R} \rightarrow T^{*}_{x}\mathbf{Q} $ as the partial derivative of the operatee at $x\in \mathbf{Q}$ with respect to the subscripted configuration variable $u$ to yield a vector in the cotangent space $T^{*}_{x}\mathbf{Q}$. The variation of the action integral is expressed by the following equation 
\small
\begin{align*}
\delta\mathcal{A}&=\delta\int_{t_0}^{t_f}\mathcal{L}dt=\int_{t_0}^{t_f}\delta\mathcal{L}dt\\
&=\int_{t_0}^{t_f}(\mathbf{D}_{\dot{o}_p}\mathcal{L}\cdot\delta\dot{o}_p+\mathbf{D}_{o_p}\mathcal{L}\cdot\delta o_p+\mathbf{D}_{\dot{r}_b}\mathcal{L}\cdot\delta\dot{r}_b\\
&+\mathbf{D}_{r_b}\mathcal{L}\cdot\delta r_b+\mathbf{D}_{\Omega_p}\mathcal{L}\cdot\delta\Omega_p+\mathbf{D}_{R_p}\mathcal{L}\cdot\delta R_p\\
&+\mathbf{D}_{\dot{q}_i}\mathcal{L}\cdot\delta\dot{q}_i+\mathbf{D}_{q_i}\mathcal{L}\cdot\delta q_i\\&+\sum_{i=1}^{3}\mathbf{D}_{\Omega_i}\mathcal{L}\cdot\delta\Omega_i+\mathbf{D}_{R_i}\mathcal{L}\cdot\delta R_i)dt
\end{align*}
\normalsize
To derive coordinate-free equations of motion, we use the exponential map to express infinitesimal variations of elements belonging to a Lie group as follows
$$\delta g =\left.\frac{d}{d\epsilon }\right|_{\epsilon=0} \textrm{exp}(\epsilon\chi)g$$ where $g$ is an element of a Lie group $G$, $\chi$ is an element of the Lie algebra $\mathfrak{g}$ of $G$ and $\textrm{exp} : \mathfrak{g}\rightarrow G$ is the exponential map(\cite{bullobook}). 

The rotation matrices belong to the Lie group $\mathop{SO}(3)$ with its Lie algebra being the space of skew-symmetric matrices.
 The variation of a rotation matrix $R$ can thus be expressed as $$\delta R=R\hat{\eta}$$ where $\eta \in \mathbb{R}^3$ is mapped to the Lie algebra element via the $\hat{\cdot}$ map. The variation of a unit vector $q \in \mathop{S}^2$ can be obtained by using the fact that any two unit vectors are uniquely related by a rotation matrix as follows
 $$q^\epsilon=\textrm{exp}(\epsilon\hat{\xi})q$$
 $$\Rightarrow \delta q= \left.\frac{d}{d\epsilon }\right|_{\epsilon=0} q^\epsilon=\hat{\xi}q=\xi\times q$$
 where $\xi \in \mathbb{R}^3$ is mapped to the Lie algebra element (of $\mathop{SO}(3)$) via the $\hat{\cdot}$ map again.
 
 Using the rotational kinematic equations and the fact that the time derivative and variational operators commute, the variations of $\dot{q}$ and $\Omega$ are obtained as $$\delta \dot{q}=\dot{\xi}\times q + \xi\times\dot{q}$$ $$\delta \Omega = \dot{\eta}+\Omega\times\eta$$ 

 Let $u_i=f_iR_ie_3$ denote  the thrust acting on the $i$th quadrotor expressed in the 
 spatial frame. Henceforth, we shall refer to our control inputs in terms of $u_i$s and $M_i$s.
 The virtual work done by the external forces (the forces and torques acting on each quadrotor) is given by $$\delta \mathcal{W}=\int_{t_o}^{t_f}\sum_{i=1}^{3}(u_i\cdot(\delta o_p +\delta R_p x_i +l_i\delta q_i) +M_i\cdot\eta_i)dt$$ For variations of trajectories with fixed end points, the Langrange- D'Alembert principle gives us \begin{equation}\label{LdA}\delta\mathcal{A}=-\delta\mathcal{W}\end{equation}Substituting the expressions of the variations in (\ref{LdA}), using integration by parts and the fact that the variations are arbitrary, the following Euler-Lagrange equations are obtained
\small
\begin{align*}
&\dfrac{d}{dt}\mathbf{D}_{\dot{o}_p}\mathcal{L}-\mathbf{D}_{o_p}\mathcal{L}=\sum_{i=1}^3u_i 
\;\;\;\; \hbox{\{Plate translation\}} \\
&\dfrac{d}{dt}\mathbf{D}_{\dot{r}_b}\mathcal{L}-\mathbf{D}_{r_b}\mathcal{L}=0
\;\;\;\; \hbox{\{Ball dynamics\}} \;\;\;\;  \\
&\dfrac{d}{dt}\mathbf{D}_{\Omega_p}\mathcal{L}+\Omega_p\times\mathbf{D}_{\Omega_p}\mathcal{L}-\mathbf{D}_{R_p}\mathcal{L}=\sum_{i=1}^3\hat{x}_iR^T_pu_i  \;\;\;\; \hbox{\{Plate orientation\}} \\
&\hat{q}_i\dfrac{d}{dt}\mathbf{D}_{\dot{q}_i}\mathcal{L}-\hat{q}_i\mathbf{D}_{q_i}\mathcal{L}=l_i\hat{q}_iu_i
 \;\;\;\; \hbox{\{$i$th Tether dynamics\}}\\
&\dfrac{d}{dt}\mathbf{D}_{\Omega_i}\mathcal{L}+\Omega_i\times\mathbf{D}_{\Omega_i}\mathcal{L}-\mathbf{D}_{R_i}\mathcal{L}=M_i \ \ \hbox{\{$i$th Quadrotor dynamics\}}
\end{align*}

\normalsize
The partial derivatives involved in the above computations are given by
\small
\begin{align*}
\mathbf{D}_{\dot{o}_p}\mathcal{L}=&m_p\dot{o}_p+\sum_{i=1}^3m_i(\dot{o}_p+R_p\hat{\Omega}_px_i+l_i\hat{\omega}_iq_i)\\
&+m_b(\dot{o}_p+R_p\hat{\Omega}_pEr_b+R_pE\dot{r}_b)\\
\mathbf{D}_{o_p}\mathcal{L}=&-m_pge_3-\sum_{i=1}^3m_ige_3-m_bge_3\\
\mathbf{D}_{\dot{r}_b}\mathcal{L}=&m_bE^TR^T_p(\dot{o}_p+R_p\hat{\Omega}_pEr_b+R_pE\dot{r}_b)\\
\mathbf{D}_{r_b}\mathcal{L}=&-m_bE^T\hat{\Omega}_pR^T_p(\dot{o}_p+R_p\hat{\Omega}_pEr_b+R_pE\dot{r}_b)-m_bgE^TR^T_pe_3\\
\mathbf{D}_{\Omega_p}\mathcal{L}=&(J_p-m_b\widehat{[Er_b]}^2-\sum_{i=1}^3m_i\hat{x}^2_i)\Omega_p+\sum_{i=1}^3m_i\hat{x}_iR^T_p(\dot{o}_p+l_i\dot{q}_i)\\
&+m_p\widehat{[Er_b]}R^T_p(\dot{o}_p+R_pE\dot{r}_b)\\
\mathbf{D}_{R_p}\mathcal{L}=&\sum_{i=1}^3m_i(\widehat{[\hat{\Omega}_px_i]}R^T_p(\dot{o}_p+l_i\dot{q}_i)-g\hat{x}_iR^T_pe_3)\\
&+m_b(\widehat{[\hat{\Omega}_pEr_b]}R^T_p(\dot{o}_p+R_pE\dot{r}_b)-g\widehat{[Er_b]}R^T_pe_3)\\
&+m_b\widehat{[E\dot{r}_b]}R^T_p(\dot{o}_p+R_p\hat{\Omega}_pEr_b)\\
\mathbf{D}_{\dot{q}_i}\mathcal{L}=&m_i(l_i\dot{o}_p+l_i^2\dot{q}_i+l_iR_p\hat{\Omega}_px_i)\\
\mathbf{D}_{q_i}\mathcal{L}=&-m_igl_ie_3\\
\mathbf{D}_{\Omega_i}\mathcal{L}=&J_i\Omega_i\\
\mathbf{D}_{R_i}\mathcal{L}=&0\\
\end{align*}
\normalsize
On substituting the above derivatives into the Euler-Lagrange equations, we obtain
\small
\begin{align}
&(m_p+\sum_{i=1}^3m_i+m_b)\ddot{o}_p+\sum_{i=1}^3m_i(R_p\hat{\Omega}^2_px_i-R_p\hat{x}_i\dot{\Omega}_p-l_i(\hat{q}_i\dot{\omega}_i-\hat{\omega}^2_iq_i))\nonumber\\
&+m_b(R_p\hat{\Omega}^2_pEr_b-R_p\widehat{[Er_b]}\dot{\Omega}_p+2R_p\hat{\Omega}_pE\dot{r}_b+R_pE\ddot{r}_b)\nonumber\\
&+(m_p+\sum_{i=1}^3m_i+m_b)ge_3=\sum_{i=1}^3u_i\\
&m_bE^T(R^T_p\ddot{o}_p+\hat{\Omega}^2_pEr_b-\widehat{[Er_b]}\dot{\Omega}_p+2\hat{\Omega}_pE\dot{r}_b+E\ddot{r}_b+R^T_pge_3)=0\\
&(J_p-m_b\widehat{[Er_b]}^2-\sum_{i=1}^3m_i\hat{x}^2_i)\dot{\Omega}_p+\hat{\Omega}_p(J_p-m_b\widehat{[Er_b]}^2-\sum_{i=1}^3m_i\hat{x}^2_i)\Omega_p\nonumber\\
&+\sum_{i=1}^3m_i\hat{x}_iR^T_p(-l_i(\hat{q}_i\dot{\omega}_i-\hat{\omega}^2_iq_i))+m_b\widehat{[Er_b]}(R^T_p\ddot{o}_p+2\hat{\Omega}_pE\dot{r}_b+E\ddot{r}_b)\nonumber\\
&+\sum_{i=1}^3m_i\hat{x}_iR^T_p\ddot{o}_p=\sum_{i=1}^3\hat{x}_iR^T_p(u_i-m_ige_3)-m_bg\widehat{[Er_b]}R^T_pe_3\\
&m_i(\hat{q}_i\ddot{o}_p+l_i\dot{\omega}_i+\hat{q}_iR_p\hat{\Omega}^2_px_i-\hat{q}_iR_p\hat{x}_i\dot{\Omega}_p)=\hat{q}_i(u_i-m_ige_3)\\
&J_i\dot{\Omega}_i+\hat{\Omega}_iJ_i\Omega_i=M_i
\end{align}
\normalsize
Note the coupling between the tethers and the plates in the first, third and fourth equation. 
To eliminate the angular accelerations of the tethers
$\dot{\omega}_i$ in the plate dynamics, we replace the expressions for $\dot{\omega}_i$ from equation (8) into (5) and (7), resulting in

\small
\begin{align}
&(m_p+\sum_{i=1}^3m_iq_iq^T_i)(\ddot{o}_p+ge_3)+\sum_{i=1}^3m_iq_iq^T_i(R_p\hat{\Omega}^2_px_i-R_p\hat{x}_i\dot{\Omega}_p)  \nonumber\\
&+m_b(\ddot{o}_p+R_p\hat{\Omega}^2_pEr_b-R_p\widehat{[Er_b]}\dot{\Omega}_p+2R_p\hat{\Omega}_pE\dot{r}_b+R_pE\ddot{r}_b+ge_3)  \nonumber\\
&-m_il_i||\omega||^2_iq_i=\sum_{i=1}^3q_iq^T_iu_i  \;\;\;\; \quad\hbox{\{Plate translation\} } \\
&m_bE^T(R^T_p\ddot{o}_p+\hat{\Omega}^2_pEr_b-\widehat{[Er_b]}\dot{\Omega}_p+2\hat{\Omega}_pE\dot{r}_b+E\ddot{r}_b+R^T_pge_3)=0  \nonumber   \\   
& \;\;\;\; \hbox{\{Ball dynamics\}}  \\
&(J_p-\sum_{i=1}^3m_i\hat{x}_iR^T_pq_iq^T_iR_p\hat{x}_i)\dot{\Omega}_p+\hat{\Omega}_pJ_p\Omega_p \nonumber\\
&+\sum_{i=1}^3m_i\hat{x}_iR^T_pq_iq^T_i(\ddot{o}_p+ge_3-l_i||\omega||^2_iq_i)\nonumber\\
&+m_b\widehat{[Er_b]}(R^T_p\ddot{o}_p+\hat{\Omega}^2_pEr_b-\widehat{[Er_b]}\dot{\Omega}_p+2\hat{\Omega}_pE\dot{r}_b+\nonumber\\&E\ddot{r}_b+R^T_pge_3)=\sum_{i=1}^3\hat{x}_iR^T_pq_iq^T_i(u_i-m_iR_p\hat{\Omega}^2_px_i)
\nonumber \\  
& \;\;\;\; \hbox{\{Plate orientation\}}\\
&m_i(\hat{q}_i\ddot{o}_p+l_i\dot{\omega}_i+\hat{q}_iR_p\hat{\Omega}^2_px_i-\hat{q}_iR_p\hat{x}_i\dot{\Omega}_p)=\hat{q}_i(u_i-m_ige_3)
\nonumber   \\
& \hbox{\{$i$th Tether dynamics\}} \\
&J_i\dot{\Omega}_i+\hat{\Omega}_iJ_i\Omega_i=M_i  \;\;\; \hbox{\{$i$th Quadrotor dynamics\}}
\end{align}
\normalsize
The term $q_iq^T_i (\cdot)$ indicates a projection operator $q_i\langle q_i,.\rangle$, where $\langle.,.\rangle$ is the inner product on $\mathbb{R}^3$. Hence, in the equations  above, terms such as
\small
\begin{align*}
& q_iq^T_i(R_p\hat{\Omega}^2_px_i-R_p\hat{x}_i\dot{\Omega}_p)\nonumber\\
&q_iq^T_iu_i \nonumber  \\
&q_iq^T_iR_p\hat{x}_i\dot{\Omega}_p  \nonumber\\
&q_iq^T_i(\ddot{o}_p+ge_3-l_i||\omega||^2_iq_i)\nonumber\\
&q_iq^T_i(u_i-m_iR_p\hat{\Omega}^2_px_i)
\end{align*}
\normalsize
indicate the projection of a quantity along the direction of the $i$th tether.
\section{Control Design}\label{cd}
The first set of inputs $\{ f_i, M_i \}$ has been transformed to $\{ u_i, M_i \}$. The $u_i$s
were the thrust vectors expressed in the spatial frame. We 
now introduce another decomposition of the $u_i$s as, along the tether and perpendicular to the tether.
Observe that the inputs controlling the translational dynamics of the plate and rotational dynamics of the plate appear as $q_iq^T_iu_i$ which is essentially the component of force $u_i$ along the $i$-th tether. 
For convenience, $q_iq^T_iu_i$ is denoted as $u^{||}_i$ and the orthogonal component $u^{\perp}_i$ is defined such that $$u_i=u^{||}_i+u^{\perp}_i$$
It can be seen that equations (10) and (12), which describe the rotational and translational dynamics of the plate, are solely affected by the $u^{||}_i$s whereas equation (13) which describes the dynamics of the tethers is solely affected by the $u^{\perp}_i$s. Thus, we adopt a procedure similar to that used in \cite{lee} where the controls $u_i$ and $M_i$ are designed in two steps-First, the quadrotors are replaced by fully actuated point masses and $u^{||}_i$s and $u^{\perp}_i$s are designed independently to meet the control objectives. Then the $M_i$s and $f_i$s are designed for the quadrotors such that the thrust $f_iR_ie_3$ equals $u_i=u^{||}_i+u^{\perp}_i$. 
\subsection{Design of Parallel components}
Before designing the parallel component of control $u^{||}_i$, we decouple the ball and plate system from the quadrotors by making the following observation.

We define $\mu_i$ to be the tension in the tether and then examine the free body diagram (figure \ref{f:sys}) to apply newton's second law for the $i$th quadrotor along $q_i$. 
\begin{figure}[H]
\centering
\includegraphics[scale=0.25]{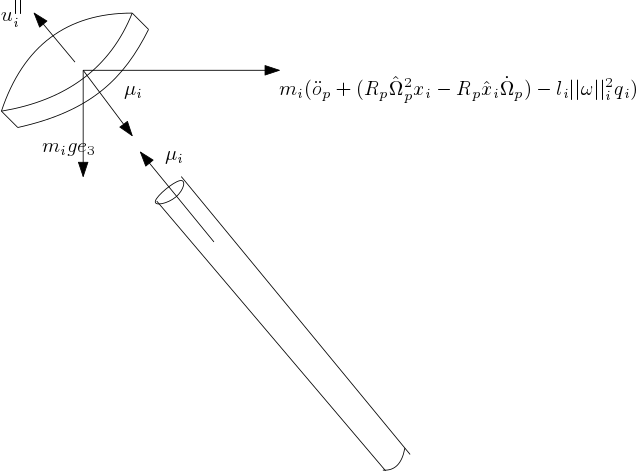}
\caption{Free body diagram of the $i$th quadrotor}
\label{f:sys}
\end{figure}
\small
\begin{align}\label{pll}
&m_iq_iq^T_i\dfrac{d^2}{dt^2}\left(o_p+R_px_i+l_iq_i\right)=u^{||}_i-\mu_i-m_iq_iq^T_ige_3\nonumber \\
&\Rightarrow m_iq_iq^T_i(\ddot{o}_p+(R_p\hat{\Omega}^2_px_i-R_p\hat{x}_i\dot{\Omega}_p)-l_i||\omega||^2_iq_i)=u^{||}_i-\mu_i-m_iq_iq^T_ige_3\nonumber\\
&\Rightarrow\mu_i=u^{||}_i-q_iq^T_im_i(\ddot{o}_p+(R_p\hat{\Omega}^2_px_i-R_p\hat{x}_i\dot{\Omega}_p)+ge_3-l_i||\omega||^2_iq_i)
\end{align}
\normalsize
where
\small$$\dfrac{d^2}{dt^2}\left(o_p+R_px_i+l_iq_i\right)=\ddot{o}_p+R_p\hat{\Omega}^2_px_i-R_p\hat{x}_i\dot{\Omega}_p-l_i||\omega||^2_iq_i$$\normalsize
 is the acceleration of the $i$th quadrotor, $u^{||}_i$ is the external force being applied parallel to $q_i$ and $-q_iq^T_im_ige_3$ is the gravitational force acting along $q_i$.\\

%
Expressing the dynamics of the ball and that of the plate (equations (10), (11) and (12)) in terms of these tensions $\mu_i$s, we obtain the dynamics of the ball and plate system completely decoupled from the quadrotors as
\small
\begin{align}
 &m_bE^T(R^T_p\ddot{o}_p+\hat{\Omega}^2_pEr_b-\widehat{[Er_b]}\dot{\Omega}_p+2\hat{\Omega}_pE\dot{r}_b+E\ddot{r}_b+R^T_pge_3)=0\\
 &m_p(\ddot{o}_p+ge_3)+\nonumber\\
 &m_b(\ddot{o}_p+R_p\hat{\Omega}^2_pEr_b-R_p\widehat{[Er_b]}\dot{\Omega}_p+2R_p\hat{\Omega}_pE\dot{r}_b+R_pE\ddot{r}_b+ge_3)=\sum_{i=1}^3\mu_i\\
  &J_p\dot{\Omega}_p+\hat{\Omega}_pJ_p\Omega_p+m_b\widehat{[Er_b]}(R^T_p\ddot{o}_p+\hat{\Omega}^2_pEr_b-\widehat{[Er_b]}\dot{\Omega}_p\nonumber\\&+2\hat{\Omega}_pE\dot{r}_b+E\ddot{r}_b+R^T_pge_3)=\sum_{i=1}^3\hat{x}_iR^T_p\mu_i
  \end{align}
  \normalsize
  We proceed to design $\mu_i$s to stabilize the attitude and position of the plate and simultaneously stabilize the ball at the centre of the plate. Once suitable $\mu_i$s have been chosen, the controls $u^{||}_i$ can be implemented by substituting for the accelerations from (16), (17) and (18) into (15).\\
 
 \subsection*{Partial Feedback Linearization}
 
The ball and plate system is described by the configuration manifold $Q_{ball-plate}=\underbrace{\mathbb{R}^2}_{Ball}\times\underbrace{\mathbb{R}^3\times\mathop{SO}(3)}_{Plate}$. This system is acted upon by a force $F=\sum_{i=1}^3\mu_i$ and a torque $\tau=\sum_{i=1}^3\hat{x}_iR^T_p\mu_i$. Thus the ball and plate system is underactuated by $2+3+3 -6=2$ degrees of freedom. We attempt to simplify the procedure to design the $\mu_i$s by employing the technique of Partial Feedback Linearization (PFL) to linearize the translational and rotational dynamics of the plate. PFL is a standard technique for addressing underactuated systems wherein a partially linearizing feedback is implemented to simplify the system dynamics before control design for meeting the desired specifications is considered (\cite{PFL}).
\begin{itemize}
\item {\bf Step 1}: Recasting the translational and attitude dynamics of the plate, and the
ball dynamics, to employ partial feedback linearization.  We proceed to rewrite the system dynamics by identifying invertible blocks in the Riemannian metric $M_{bp}$ of the ball and plate system as
$$M_{bp}=\begin{bmatrix}
m_b\mathbb{I}_2 & m_bE^TR^T_p  &-m_bE^T\widehat{[Er_b]}\\
m_bR_pE & (m_p+m_b)\mathbb{I}_3 & -m_bR_p\widehat{[Er_b]}\\
m_b\widehat{[Er_b]}E & m_b\widehat{[Er_b]}R^T_p & J_p-m_b\widehat{[Er_b]}^2
\end{bmatrix}$$
Defining 
\begin{align*}
&M_{11}=m_b\mathbb{I}_2\\ 
&M_{12}=[m_bE^TR^T_p\ -m_bE^T\widehat{[Er_b]}]\\ 
&M_{22}=\begin{bmatrix}(m_p+m_b)\mathbb{I}_3 & -m_bR_p\widehat{[Er_b]}\\
m_b\widehat{[Er_b]}R^T_p & J_p-m_b\widehat{[Er_b]}^2\end{bmatrix},
\end{align*}
we rewrite equations (16), (17) and (18) as 
\small
\begin{align}
M_{11}\ddot{r}_b+M_{12}[\ddot{o}^T_p\ \dot{\Omega}^T_p]^T + N_1=0\\
M^T_{12}\ddot{r}_b+M_{22}[\ddot{o}^T_p\ \dot{\Omega}^T_p]^T +N_2=[F^T\ \tau^T]^T
\end{align}
\normalsize
where 
\small
\begin{align*}
&N_1=m_bE^T(\hat{\Omega}_p^2Er_b+2\hat{\Omega}_pE\dot{r}_b+R^T_pge_3) \\
&N_2=\begin{bmatrix}m_pge_3 +m_bR_p\hat{\Omega}_p^2Er_b+2R_p\hat{\Omega}_pE\dot{r}_b+m_bge_3\\\hat{\Omega}_pJ_p\Omega_p+m_b\widehat{[Er_b]}(\hat{\Omega}_p^2Er_b+2\hat{\Omega}_pE\dot{r}_b+R^T_pge_3)\end{bmatrix}
\end{align*}
\normalsize
\item {\bf Step 2}: Expressing the $r_b$ dynamics in terms of the $o_p$ and $\Omega_p$ 
dynamics, and then 
cancelling the nonlinearities in the $o_p$ and $\Omega_p$ equations by defining new inputs $U_1$ and $U_2$:
Noting that $M_{11}$ is invertible, $\ddot{r}_b$ is substituted from (19) into (20) and the linearizing feedback given by $$\begin{bmatrix}F\\ \tau\end{bmatrix}=N_2-M^T_{12}M^{-1}_{11}N_1+(M_{22}-M^T_{12}M^{-1}_{11}M_{12})\begin{bmatrix}U_1\\U_2\end{bmatrix}$$ where $U_1$ and $U_2$ are the new inputs to the system, is substituted into (19) and (20) to obtain the partially linearized equations

\begin{align*}
\ddot{r}_b&=- M_{11}^{-1}N_1-M_{11}^{-1}M_{12}[U_1\:\:\:U_2]^T \\
\ddot{o}_p&=U_1\\
\dot{\Omega}_p&=U_2
\end{align*}
\item {\bf Step 3}: Now our objective is to stabilize the rotational dynamics of the plate, and then the position of the ball on the plate, and lastly, the translational dynamics of the plate.
The inputs $U_1$ and $U_2$ are coupled into the dynamics of the ball via the matrix $M_{11}^{-1}M_{12}=[E^TR^T_p\ -E^T\widehat{[Er_b]}]$. $E^TR^T_p$ has a constant rank of 2, $R_pe_3$ spans its null space and $E^TR^T_pR_pE=\mathbf{I}_{2}$. Using these facts, $U_1$ is chosen as 
\begin{equation}U_1=R_pe_3e^T_3(-k_5\dot{o}_p-k_6o_p)+\frac{1}{m_b}R_pE(-N_1+M_{11}(k_4r_b+k_3\dot{r}_b))\end{equation}
To stabilize the attitude of the plate, $U_2$ is chosen as \begin{equation}U_2=-k_2\eta-k_1\Omega_p\end{equation} where $\eta$ is the gradient of $\Psi(R)=\frac{1}{2}\mathrm{trace}{\left(I_{3\times 3}-R_p\right)}$ (see \cite{leetrack} for details on exponential attitude stabilization).
The closed loop dynamics are described by the following equations
\begin{tcolorbox}
\small
\begin{align}
\ddot{r}_b&=-k_4r_b-k_3\dot{r}_b+E^T\widehat{[Er_b]}U_2\\
\ddot{o}_p&=R_pe_3e^T_3(-k_5\dot{o}_p-k_6o_p)\nonumber\\&+\frac{1}{m_b}R_pE(-N_1+M_{11}(k_4r_b+k_3\dot{r}_b))\\
\ddot{\Omega}_p&=-k_2\eta-k_1\Omega_p
\end{align}
\normalsize
\end{tcolorbox}
Examine the first and the third equation. Choosing appropriate gains $k_2$ and $k_1$  ensures that the attitude gets stabilized exponentially fast, which in turn ensures that $U_2$ approaches 0 exponentially fast.
This leads to the last term in equation (23) going to zero. An appropriate choice of $k_4$ and $k_3$ esnures that $r_b$ 
asymptotically goes to zero. When both $r_b$ and the attitude have been stabilized, the dynamics of the z coordinate of the CoM of the plate are given by $$e^T_3\ddot{o}_p=e^T_3(-k_5\dot{o}_p-k_6o_p)$$ which, for appropriate choices of $k_5$ and $k_6$, stabilizes the z coordinate of the CoM of the plate as well.
\end{itemize}
\begin{thm}\label{stabilityproof}
Consider the closed loop system described by equations (23), (24) and (25). Then there exist positive scalars $k_1,\ k_2,\ k_3,\ k_4,\ k_5,\ k_6$ such that the state $(r_b,\dot{r}_b,\ e^T_3o_p,e^T_3\dot{o}_p,R_p,\Omega_p)=([0\ 0]^T,[0\ 0]^T,0,0,\mathbf{I}_3,[0\ 0\ 0]^T)$ 
is asymptotically stable.\\
\end{thm}

\subsubsection*{Proof}
To prove the stability of the state $(r_b,\dot{r}_b,\ e^T_3o_p,e^T_3\dot{o}_p,R_p,\Omega_p)=([0\ 0]^T,[0\ 0]^T,0,0,\mathbf{I}_3,[0\ 0\ 0]^T)$ of the system described by equations (23), (24) and (25), consider the following Lyapunov function candidate
\small
\begin{align*}V=\frac{k_4+c_1k_3}{2}||r_b||^2+c_1r^T_b\dot{r}_b+\frac{1}{2}||\dot{r}_b||^2+(k_2+c_2k_1)\Psi\\
+c_2\eta^T\Omega_p+\frac{1}{2}||\Omega_p||^2+\frac{k_6}{2}||o^T_pe_3||^2+\frac{1}{2}||\dot{o}^T_pe_3||^2
\end{align*}
\normalsize

which is positive definite for small values of $c_1$ and $c_2$. The time derivative of the candidate function along the system trajectories is given by
\small
\begin{align}
\dot{V}&=k_4r^T_b\dot{r}_b+\dot{r}^T_b\ddot{r}_b+c_1||\dot{r}_b||^2+c_1r^T_b\ddot{r}_b+k_2\eta^T\Omega_p+\Omega^T_p\dot{\Omega}_p+\nonumber\\&c_2\dot{\eta}^T\Omega_p+c_2\eta^T\dot{\Omega}_p+k_6(o_p^Te_3)(\dot{o}^T_pe_3)+(\dot{o}^T_pe_3)(\ddot{o}^T_pe_3)\nonumber\\
&\leq-(k_1-c_2)||\Omega_p||^2-(k_3-c_2)||\dot{r}_b||^2-c_1k_4||r_b||^2-c_2k_2||\eta||^2-\nonumber\\&k_5||\dot{o}^T_pe_3||^2+\dot{r}^T_bE^T\widehat{[Er_b]}(-k_2\eta_E-k_1\Omega_p)+\nonumber\\
&\dot{o}^T_pe_3e^T_3((R_p-I)e_3e^T_3(-k_5\dot{o}_p-k_6o_p)+\nonumber\\&\frac{1}{m_b}R_pE(-N_1+M_{11}(k_4r_b+k_3\dot{r}_b))
\end{align}
\normalsize
The cubic terms in the derivative can be bounded as follows 
\small
\begin{align*}&\dot{r}^T_bE^T\widehat{[Er_b]}(-k_2\eta_E-k_1\Omega_p)\leq ||\dot{r}_b||||r_b||(||k_2||\eta||+k_1||\Omega_p||)\\
&\dot{o}^T_pe_3e^T_3(R_p-I)e_3e^T_3(-k_5\dot{o}_p-k_6o_p)\leq 2||o^T_pe_3||(k_5||\dot{o}^T_pe_3||+k_6||o^T_pe_3||)\\
&\dot{o}^T_pe_3e^T_3\frac{1}{m_b}R_pE(-N_1+M_{11}(k_4r_b+k_3\dot{r}_b))\leq ||\dot{o}^T_pe_3||(||\Omega_p||^2||r_b||+\\&2||\Omega_p||||\dot{r}_b||+k_4||r_b||+k_3||\dot{r}_b||)+g||\dot{o}^T_pe_3|||(-1+||e_3^TR^T_pe_3||^2)|
\end{align*}
\normalsize
Substituting the above bounds into (26), the derivative of the candidate function is bounded above by

\begin{align}
\dot{V}\leq z^T\mathcal{W}z+\mathcal{O}
\end{align}
where \small \begin{align*}
z&=[||r_b||\ ||\dot{r}_b||\ ||\eta||\ ||\Omega_p||\ ||o_p^Te_3||\ ||\dot{o}^T_pe_3||]^T\\
\mathcal{W}&=\begin{bmatrix}
				-c_1k_4 & 0 & 0 & 0 & 0 & \frac{k_4}{2}\\
				0 & -k_3+c_1 & 0 &0 & 0 & \frac{k_3}{2}\\
				0 & 0 & -c_2k_2 & 0 & 0 & 0 \\
				0 & 0 & 0 & -k_1+c_2 & 0 & 0\\
				0 & 0 & 0 & 0 & 2k_6 & k_5\\
				\frac{k_4}{2} & \frac{k_3}{2} & 0 & 0 & k_5 & -k_5 
				\end{bmatrix}\\
\mathcal{O}&= ||\dot{r}_b||||r_b||(k_2|\eta||+k_1||\Omega_p||) + ||\dot{o}^T_pe_3||(||\Omega_p||^2||r_b|| +\\& 2||\Omega_p||||\dot{r}_b||+g|(-1+||e_3^TR^T_pe_3||^2)|)
\end{align*}
\normalsize
To bound the cubic and quartic terms in $\mathcal{O}$, we obtain bounds on $\eta$ and $\Omega_p$ as follows.\\\\
Consider the differentiable function $$V_2=\frac{1}{2}||\Omega_p||^2+ c_{0}\eta^T\Omega_p+(k_2+c_0k_1)\Psi$$ defined on the state space describing the plate's attitude, $(R_p,\ \Omega_p)$. This function  can be shown to be positive definite on this state space if $c_0$ is small. The time derivative of the function is given by \small\begin{align*}
\dot{V}_2&=- k_1||\Omega_p||^2-c_0 k_2||\eta||^2+c_0\dot{\eta}^T\Omega_p\\
&\leq(-k_1+c_0)||\Omega_p||^2-c_0k_2||\eta||^2
\end{align*}\normalsize
where the second inequality follows from that fact that $||\dot{\eta}||\leq ||\Omega_p||$. For $k_1>c_0$, the time derivative of $V_2$ is bounded above by a negative definite quantity on the state space $(R_p,\ \Omega_p)$ and is thus, negative definite on this space as well. This implies that $V_2$ decays to zero exponentially and is bounded above by its initial value $V_2(0)$. Since $V_2$ is positive definite on the considered space, we have that $$||\Omega_p|| \leq C_1(V_2(0))\quad ||\eta||\leq C_2(V_2(0))$$ where $C_1$ and $C_2$ are some constants in terms of $V_2(0)$. These bounds can be used to express the cubic and quartic terms as quadratic terms to bound the derivative of $V$ in (27) as 
\begin{align}
\dot{V}\leq z^T\mathcal{W'}z+||\dot{o}^T_pe_3||g|(-1+||e_3^TR^T_pe_3||^2)|
\end{align}
where
\small
$$\mathcal{W'}=\begin{bmatrix}
				-c_1k_4 & \frac{k_2C_2+k_1C_1}{2} & 0 & 0 & 0 & \frac{k_4}{2}+\frac{C^2_1}{2}\\
				\frac{k_2C_2+k_1C_1}{2} & -k_3+c_1 & 0 &0 & 0 & \frac{k_3}{2}+\frac{C^2_2}{2}\\
				0 & 0 & -c_2k_2 & 0 & 0 & 0 \\
				0 & 0 & 0 & -k_1+c_2 & 0 & 0\\
				0 & 0 & 0 & 0 & 2k_6 & k_5\\
				\frac{k_4}{2}+\frac{C^2_1}{2} & \frac{k_3}{2}+\frac{C^2_2}{2} & 0 & 0 & k_5 & -k_5 
				\end{bmatrix}$$
\normalsize
Note that the constants $C_1$ and $C_2$ appear in the matrix $\mathcal{W'}$. The right hand side of inequality (28) can be shown to be negative definite when $c_1,\ c_2$ are small, $k_3>c_1$, $k_1>c_2$, $k_1,\ k_2$ are large and $||\dot{o}^T_pe_3||>\frac{g|(-1+||e_3^TR^T_pe_3||^2)|}{k_5}$. Note that since $V_2$ decays to $0$ exponentially, we have $\eta$ decaying to zero exponentially as well. This implies that $R_p$ approaches identity and the condition on $\dot{o}^T_pe_3$ becomes $||\dot{o}^T_pe_3||>0$. Thus, under these conditions we have $$\dot{V}<0$$ and the desired state is asymptotically stable using Lyapunov's direct stability theorem (\cite{khalil}).
\\
$\Box$.

\subsection*{The internal dynamics}
To investigate the behaviour of the unstabilized states (the internal dynamics), we analyze the translational dynamics of the plate in the x-y plane
\begin{align}
\ddot{o}_{p1}=e^T_1\ddot{o}_p=e^T_1U_1\\
\ddot{o}_{p2}=e^T_2\ddot{o}_p=e^T_2U_1
\end{align}
When the ball positon $r_b$, the attitude $R_p$ and the $z$-coordinate ($o_{p3}$) of the plate get stabilized, we have $U_1 \rightarrow 0$ and thus $\ddot{o}_{p1}  \rightarrow 0$ and
$\ddot{o}_{p2}  \rightarrow 0$. 
\begin{remark}
We seek to establish a result converse to that stated by Barbalat's lemma, i.e, given $\lim_{t\rightarrow\infty}\ddot{o}_{pi}= 0$, we seek to prove $\lim_{t\rightarrow\infty}\dot{o}_{pi}=c_i$ for some constants $c_i,\ i=1,2$. 
\end{remark}
\begin{claim}
The translational velocities of the pate in the $x$ and $y$ direction satisfy
\begin{align*}
\lim_{t\rightarrow\infty}\dot{o}_{p1}(t)= c_1 \;\;\;\; \lim_{t\rightarrow\infty}\dot{o}_{p2}(t)= c_2
\end{align*} 
where $c_1$ and $c_2$ are finite constants.
\end{claim}
To prove the above claim, we employ a result from input-to-state stability, which is now reproduced from
 \cite{sontag}.
 %
 
 %

\begin{thm}\cite{sontag}\label{iss}
Internally stable linear systems $\dot{x}=Ax+Bu$ are \textit{Input to State Stable} (ISS) and the state satisfies the following ISS estimate
$$|x(t)|\leq ||e^{tA}|||x(0)| + ||B||\int_{0}^{\infty}||e^{sA}||ds||u||_{\infty}$$
\end{thm}

\subsubsection*{Proof of claim 1}
Since the velocities and accelerations are continuous, we prove this by establishing integrability of the accelerations. It suffices to show that the accelerations are composed of exponentially decaying terms.

From equations (29) and (30), we need to examine how the first and second components of signal $$U_1=R_pe_3e^T_3(-k_5\dot{o}_p-k_6o_p)+\frac{1}{m_b}R_pE(-N_1+M_{11}(k_4r_b+k_3\dot{r}_b))$$ behaves.
We have already shown that the attitude of the plate stabilizes ($R_p\rightarrow I$, $\Omega_p\rightarrow \mathbf{0}$) exponentially. This implies that $e^T_1R_pe_3e^T_3(-k_5\dot{o}_p-k_6o_p)$ and $e^T_2R_pe_3e^T_3(-k_5\dot{o}_p-k_6o_p)$, i.e, the first and second components of the first term of $U_1$ decay exponentially fast. To show the exponential decay of the second term, we observe the dynamics of $r_b$.
$$\ddot{r}_b=-k_4r_b-k_3\dot{r}_b+E^T\widehat{[Er_b]}U_2$$
Note that we have $U_2=[\beta_1\ \beta_2\ \beta_3]^Te^{-\gamma t}$ for some $\beta_i \in \mathbb{R}, \gamma>0$ due to the exponentially decaying attitude dynamics. Then the above dynamics can be written as
\begin{align*}
\ddot{r}_{b1}+k_4r_{b1}+k_3\dot{r}_{b1}&=\beta_3 e^{-\gamma t} r_{b2}\\
\ddot{r}_{b2}+k_4r_{b2}+k_3\dot{r}_{b2}&=-\beta_3 e^{-\gamma t} r_{b1}
\end{align*} 
In state space, the above equations are given by
\begin{align}\label{rb}
\mathbf{\dot{X}}=\begin{bmatrix}0 & 1 & 0 & 0\\-k_4& -k_3 & 0 & 0\\0&0&0&1\\0&0&-k_4&-k_3\end{bmatrix}\mathbf{X}+\beta_3e^{-\gamma t}\begin{bmatrix}0&0&0&0\\0&0&1&0\\0&0&0&0\\-1&0&0&0\end{bmatrix}\mathbf{X}\end{align} where $\mathbf{X}=[r_{b1}\  \dot{r}_{b1}\ r_{b2}\ \dot{r}_{b2}]^T$.\\
Note that 
\begin{align*}
A=\begin{bmatrix}0 & 1 & 0 & 0\\-k_4& -k_3 & 0 & 0\\0&0&0&1\\0&0&-k_4&-k_3\end{bmatrix}
\end{align*}
is Hurwitz. Let 
\begin{align*}
B=\beta_3 e^{-\gamma t}  \;\;\; \hbox{and} \;\;\;  u=\begin{bmatrix}0&0&0&0\\0&0&1&0\\0&0&0&0\\-1&0&0&0\end{bmatrix}\mathbf{X}
\end{align*}

%
We have shown that $r_b(t) \rightarrow 0$ asymptotically in the proof of theorem \ref{stabilityproof}. Moreover, $r_b(t)$ is a continuous function of $t$ because the inputs $U_1$ and $U_2$ are continuous. Thus $r_b$ is bounded and consequently $u\in L^{\infty}$ in system (\ref{rb}). 
We now invoke theorem \ref{iss} to conclude that system (\ref{rb}) is ISS and the state $X(t)$ satisfies the following inequality  
$$|X(t)|\leq ||e^{tA}|||X(0)| +|\beta_3|e^{-\gamma t}(\int_{0}^{\infty}||e^{sA}||ds)||u||_{\infty} \;\;\;\; \forall t \geq 0 $$
%

Thus we have established that
\begin{align}
||k_4r_b+k_3\dot{r}_b||&\leq \alpha_r e^{-\gamma_r t} \label{eq:rbBnd}
\end{align}
for some $\alpha_r, \gamma_r >0$. Now we examine 
 $$N_1=m_bE^T(\hat{\Omega}_p^2Er_b+2\hat{\Omega}_pE\dot{r}_b+R^T_pge_3)$$ 
 Observe that since $R_p\rightarrow I$ exponentially, the last term $E^TR^T_p[0\ 0\ g]^T\rightarrow \mathbf{0}$ exponentially. The first and second terms of $N_1$ can also be shown to bounded above by exponentially decaying terms because of the exponentially stabilized attitude dynamics and inequality (\ref{eq:rbBnd}). To consolidate the above arguments, we have successfully shown that $e^T_1U_1$ and $e_2^TU_1$ are bounded above by exponentially decaying terms and thus the accelerations $\ddot{o}_{p1}$ and $\ddot{o}_{p2}$ are integrable. Furthermore, this helps us establish that the velocities $\dot{o}_{p1}$ and $\dot{o}_{p2}$ converge to a limit as follows.
\begin{align*}
\dot{o}_{pi}(t)-\dot{o}_{pi}(0)&=\int^{t}_0\ddot{o}_{pi}(s)ds\\
\Rightarrow\lim_{t\rightarrow\infty}\dot{o}_{pi}(t)&=\dot{o}_{pi}(0)+\lim_{t\rightarrow\infty}\int^{t}_0\ddot{o}_{pi}(s)ds=c_i
\end{align*} 
Owing to the above result and the continuity of the velocities, we can conclude that $sup_{t\rightarrow\infty}\dot{o}_{pi}(t)\ i=1,2$ exist and bound the respective velocities from above. \\
$\Box$
\begin{remark}
Note that the Riemannian metric $M_{bp}$ and potential energy of the ball and plate system, are invariant to flows along the vector fields $\frac{\partial}{\partial o_{p1}}, \frac{\partial}{\partial o_{p2}} \in TQ_{ball-plate}$. These vector fields are in fact infinitesimal symmetries(\cite{bullore})
\end{remark}
\begin{remark}
 Traditional approaches to stabilization of underactuated systems after applying PFL involve stabilizing either the actuated subsystem (dynamics of plate) with the unactuated subsystem (dynamics of ball) constituting the internal dynamics or vice-versa. Both of these approaches yield non-minimum phase internal dynamics and present solutions that are not practically as viable as ours.\\
 \end{remark}
 %
 \subsubsection*{Procedure to obtain $u^{||}_i$}
\begin{enumerate} 
\item The designed $U_1$ and $U_2$ are mapped to $F$ and $\tau$ (the net force and torque acting on the plate) by the transformation
$$\begin{bmatrix}F\\ \tau\end{bmatrix}=N_2-M^T_{12}M^{-1}_{11}N_1+(M_{22}-M^T_{12}M^{-1}_{11}M_{12})\begin{bmatrix}U_1\\U_2\end{bmatrix}$$
\item The $\mu_i$s are obtained from $F$ and $\tau$ by solving the following set of linear equations
\begin{equation*}\label{forces}
\underbrace{\begin{bmatrix}  I & I & I\\ \hat{x}_1 & \hat{x}_2 & \hat{x}_3 \end{bmatrix}}_{A}\begin{bmatrix}R_p^T\mu_1\\R_p^T\mu_2\\R_p^T\mu_3\end{bmatrix}=\begin{bmatrix}R^T_pF\\\tau\end{bmatrix}
\end{equation*}
\item Matrix $A$ has a full row rank if the vectors $x_1,\ x_2$ and $x_3$ are coplanar but not collinear. Thus, the minimum norm solution for the $\mu_i$s is given by $$\begin{bmatrix}\mu_1\\\mu_2\\\mu_3\end{bmatrix}=\textrm{diag}(R_p,R_p,R_p)A^T(AA^T)^{-1}\begin{bmatrix}R^T_pF\\\tau\end{bmatrix}$$
\item The $u^{||}_i$s are obtained from equation (17) (restated here for convenience) by substituting $\ddot{o}_p=U_1$ and $\dot{\Omega}_p=U_2$
$$u^{||}_i-q_iq^T_im_i(\ddot{o}_p+(R_p\hat{\Omega}^2_px_i-R_p\hat{x}_i\dot{\Omega}_p)+ge_3-l_i||\omega||^2_iq_i)=\mu_i$$
\end{enumerate}
\subsection{Design of Perpendicular components}
The perpendicular component $u^{\perp}_i$ is chosen such that the tether is aligned in the direction $q_{id}=\frac{\mu_i}{||\mu_i||}$. Equation (13) describes the tether dynamics and is rewritten here for convenience
\small
\begin{align*}
m_i(\hat{q}_i\ddot{o}_p+l_i\dot{\omega}_i+\hat{q}_iR_p\hat{\Omega}^2_px_i-\hat{q}_iR_p\hat{x}_i\dot{\Omega}_p)=\hat{q}_i(u^{\perp}_i-m_ige_3)
\end{align*}
\normalsize
Grouping the coupled acceleration terms, we define \small$$a_i=\ddot{o}_p+R_p\hat{\Omega}^2_px_i-R_p\hat{x}_i\dot{\Omega}_p+ge_3$$\normalsize
and rewrite equation (13) as 
\small
\begin{align*}
\frac{1}{l_i}\hat{q}_ia_i+\dot{\omega}_i=\frac{1}{m_il_i}\hat{q}_iu^{\perp}_i
\end{align*}
\normalsize
We refer to \cite{lee} to solve the tracking problem by choosing the control
\small
\begin{align}
u^{\perp}_i=-m_i\hat{q}^2_ia_i +m_il_i\hat{q}_i(k_7e_{q_i}+k_8e_{\omega_i}+(q^T_i\omega_{id})\dot{q}_i+\hat{q}^2_i\dot{\omega}_d)
\end{align}
\normalsize
where $\omega_{id}=\hat{q}_{id}\dot{q}_{id}$, $e_{q_i}=\hat{q}_{id}q_i$ and $e_{\omega_i}=\omega_i+\hat{q}_i\omega_{id}$. Note that the acceleration terms in $a_i$ are substituted in terms of the $\mu_i$s for implementation. The closed loop dynamics is given by 
\small
\begin{align}
\dot{\omega}_i=-k_7e_{q_i}-k_8e_{\omega_i}-(q^T_i\omega_{id})\dot{q}_i-\hat{q}^2_i\dot{\omega}_{id}
\end{align} 
\normalsize

\subsection{Design of Quadrotor inputs}
The total control $f_iR_ie_3=u_i=u^{||}_i+u^{\perp}_i$ is to be generated by the $i$th quadrotor using the inputs $f_i$ and $M_i$. Again, this problem has been solved in \cite{lee} and we refer to their approach. A backstepping-like controller is used so that $M_i$ orients the yaw axis of the quadrotor along that of $u_i$, i.e., $R_ie_3$ tracks $\frac{u_i}{|u_i|}$.  The first two rows of the desired $R_i$ are obtained by considering some smooth $b_{1i}(t)\in \mathop{S}^2$ which is used to form a right handed coordinate system along with $b_{3i}=\frac{u_i}{|u_i|}$ by defining the following desired attitude of the quadrotor.
$$R_{id}=\begin{bmatrix}\frac{-\hat{b}_{3i}^2b_{1i}}{||\hat{b}_{3i}^2b_{1i}||} & \frac{\hat{b}_{3i}b_{1i}}{||\hat{b}_{31}b_{1i}||} & b_{3i}  \end{bmatrix}$$
Tracking errors for the attitude and angular velocity of the $i$th quadrotor are defined as 
\begin{align}
e_{R_i}=\frac{1}{2}(R^T_{id}R_{i}-R^T_{i}R_{id})^{\vee},\quad e_{\Omega_i}=\Omega_i-R^T_iR_{id}\Omega_{id}
\end{align}
where the map $\cdot^\vee$ is the inverse of the $\hat{\cdot}$ map and the desired angular velocity is obtained from the attitude kinematics as $\Omega_{id}=(R^T_{id}\dot{R}_{id})^\vee$.
With this, the thrust and moment of the quadrotor are chosen as 
\begin{align}
&f_i=||u_i||\\
&M_i=-\frac{k_{R}}{\epsilon^2}e_{R_i}-\frac{k_\Omega}{\epsilon}e_{\Omega_i}+\Omega_i\times J_i\Omega_i\nonumber\\
&-J_i(\hat{\Omega}_iR^T_iR_{id}\Omega_{id}-R^T_iR_{id}\dot{\Omega}_{id})
\end{align}
for some positive constants $\epsilon,\ k_{R}$ and $k_{\Omega}$.
\\

\begin{thm}
Consider the full dynamic model given by (10)-(14). For a desired direction of the first body-fixed axes $b_{1i}\quad i=1,2,3$, control inputs (34) and (35), there exists some $\epsilon^*>0$, such that for all $\epsilon<\epsilon^*$, the zero equilibrium of the tracking errors of the quadrotors $(e_{R_i},\ e_{\Omega_i})=(\mathbf{0},\ \mathbf{0})$ is exponentially stable and the state $(r_b,\dot{r}_b,\ e^T_3o_p,e^T_3\dot{o}_p,R_p,\Omega_p)=([0\ 0]^T,[0\ 0]^T,0,0,\mathbf{I}_3,[0\ 0\ 0]^T)$ is asymptotically stable and the velocities $\dot{o}_{p1}$ and $\dot{o}_{p2}$ remain bounded.\\
\end{thm}
\subsubsection*{Proof}
Let $\bar{e}_{R_i}=\frac{1}{\epsilon}e_{R_i}$. Before deriving the attitude error dynamics of the quadrotors, we make note of the following properties of the $\hat{.}$ map $$R\hat{x}R^T=\widehat{Rx}$$ 
$$\hat{x}A+A^T\hat{x}=(\{\textrm{tr}(A)I-A\}x\hat{) }$$
To derive the attitude error dynamics, we differentiate the equations in (33) to obtain the attitude tracking error dynamics as 
\small
\begin{align}
\epsilon\dot{\bar{e}}_{R_i}&=\frac{1}{2}(R^T_{id}R_i(\hat{\Omega}_i-R^T_iR_{id}\hat{\Omega}_{id}R^T_{id}R_i)+(\hat{\Omega}_i-R^T_iR_{id}\hat{\Omega}_{id}R^T_{id}R_i)R^T_iR_{id})^{\vee}\nonumber\\
\Rightarrow \epsilon\dot{\bar{e}}_{R_i}&=\frac{1}{2}(R^T_{id}R_i\hat{e}_{\Omega_i}+\hat{e}_{\Omega_i}R^T_iR_{id})^\vee=\frac{1}{2}(\textrm{tr}[R^T_iR_{id}]I-R^T_iR_{id})e_{\Omega_i}
\end{align}
\normalsize
and the dynamics of the tracking error of the angular velocity as 
\small
 \begin{align}
 \epsilon\dot{e}_{\Omega_i}&=\dot{\Omega}_i+\hat{\Omega}_iR^T_iR_{id}\Omega_{id}-R^T_iR_{id}\dot{\Omega}_{id}\nonumber\\
\Rightarrow\epsilon\dot{e}_{\Omega_i}&=J^{-1}_i(-k_{R}\bar{e}_{R_i}-k_{\Omega}e_{\Omega_i})
\end{align}
\normalsize
The system model is described by equations (36) and (37) (called \textit{boundary layer equations}) and equations (10)-(13).
As $\epsilon \rightarrow 0$, the error dynamics are described by $$0=\frac{1}{2}(\textrm{tr}[R^T_iR_{id}]I-R^T_iR_{id})e_{\Omega_i}$$ $$0=J^{-1}_i(-k_{R}\bar{e}_{R_i}-k_{\Omega}e_{\Omega_i})$$ the solution to which is an isolated root $(e_{R_i},\ e_{\Omega_i})=(\mathbf{0},\ \mathbf{0})$. When $(e_{R_i},\ e_{\Omega_i})=(\mathbf{0},\ \mathbf{0})$, the force on the $i$th tether is given by $u_i$ and thus, the system dynamics are described by the equations (23), (24), (25) and (32), which are the desired closed loop dynamics of the ball and plate system. These equations are called the \textit{reduced} dynamics.\\
The zero equilibrium of the error dynamics described by (35) and (36) has been shown to be exponentially stable in \cite{lee}.\\
For the reduced system, it has been shown that the state $(r_b,\dot{r}_b,\ e^T_3o_p,e^T_3\dot{o}_p,R_p,\Omega_p)=([0\ 0]^T,[0\ 0]^T,0,0,\mathbf{I}_3,[0\ 0\ 0]^T)$ is asymptotically stable, the velocities $\dot{o}_{p1}$ and $\dot{o}_{p2}$ remain bounded and that $(e_{q_i},\ e_{\omega_i})=(\mathbf{0},\ \mathbf{0})$ is exponentially stable as well.\\Then according to Tikhonov's theorem (\cite{khalil}), there exists some $\epsilon^*>0$, such that for all $\epsilon<\epsilon^*$, the hypothesis holds true. $\blacksquare$\\

\section{Simulations}\label{sim}
For simulating the behaviour of the system under the action of the proposed control laws, the following system parameters are chosen
\begin{align*}
&m_p=0.75 \ m_b=0.1\\
&J_p=\begin{bmatrix}0.006&0&0\\0&0.008&0\\0&0&0.012\end{bmatrix}
\end{align*}
Furthermore, prior to employing a numerical technique to obtain the trajectories described by equations (10)-(13), the system states are initialised to the following values.
\begin{align*}
&r_b(0)=[1\ 1]^T\quad \dot{r}_b(0)=[0.5\ 0.5]^T\\
&q_1(0)=[0\ 0\ 1]^T \ \omega_1(0)=[0\ 0\ 0]^T\\ &q_2(0)=[-0.5126\ 0.0854\ 0.8544]^T\  \omega_2(0)=[0\ 0\ 0]^T\\ &q_3(0)=[-0.5126\ 0.0854\ 0.8544]^T\ \omega_3(0)=[0\ 0\ 0]^T\\
&R_p(0)=\begin{bmatrix}
1&0&0\\0&0&-1\\0&1&0
\end{bmatrix}\ \Omega_p(0)=[1\ 1\ 2]^T
\end{align*}
All quantities are expressed in their respective SI units.
\begin{figure}[H] 
\centering
\begin{subfigure}{0.2\textwidth}
\includegraphics[width=\linewidth]{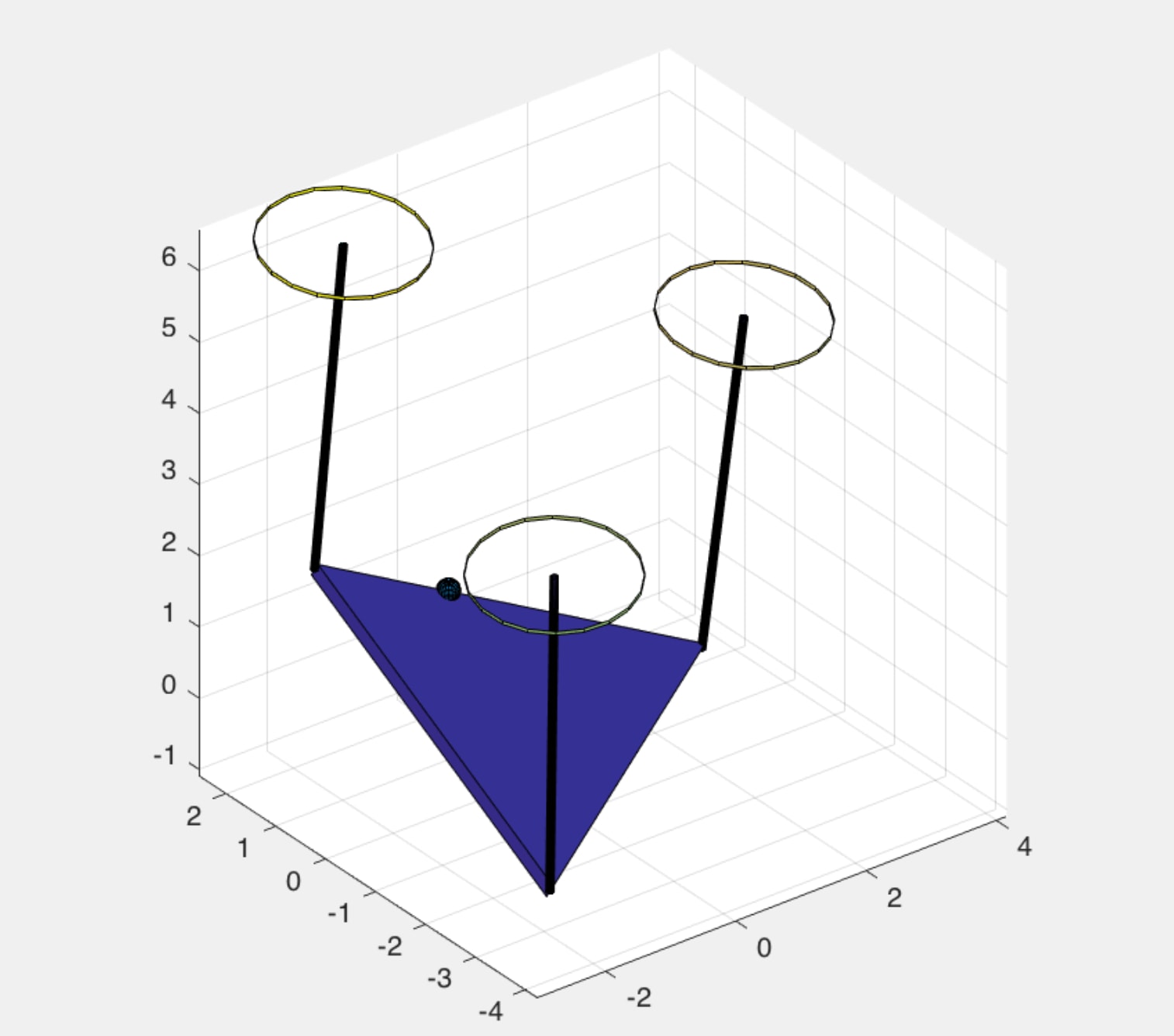}
\caption{}
\end{subfigure}
\begin{subfigure}{0.2\textwidth}
\includegraphics[width=\linewidth]{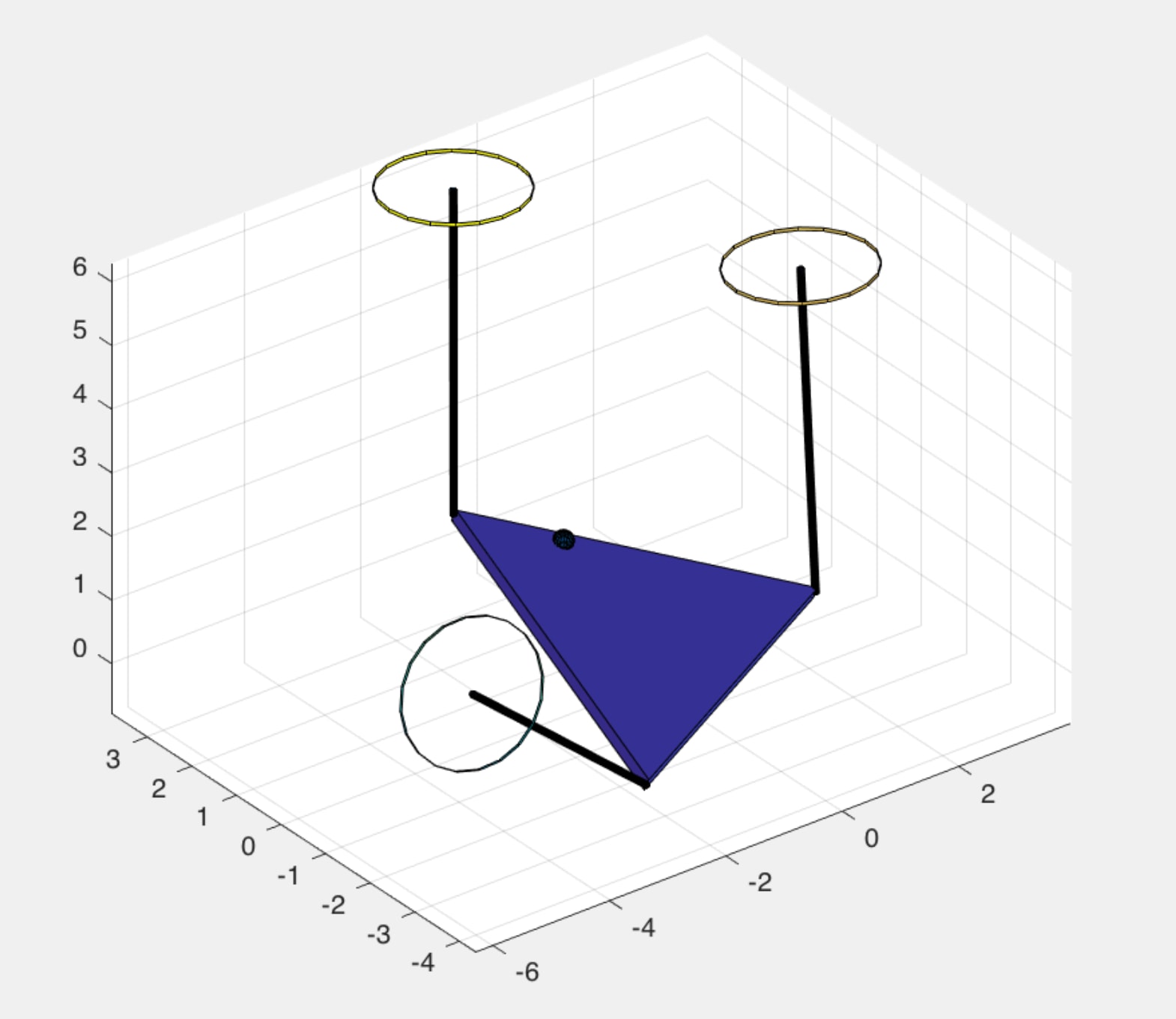}
\caption{}
\end{subfigure}

\medskip
\begin{subfigure}{0.2\textwidth}
\includegraphics[width=\linewidth]{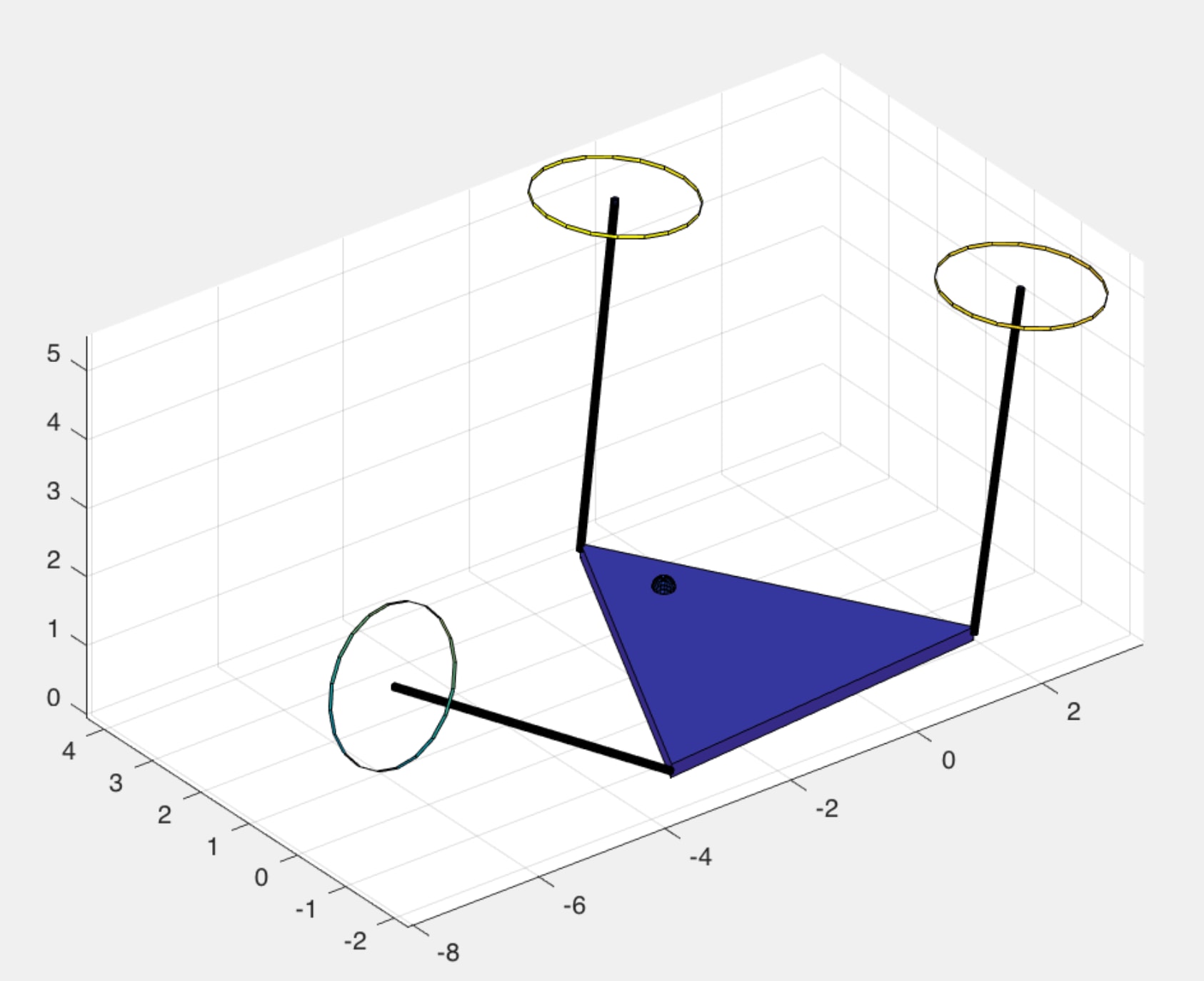}
\caption{}
\end{subfigure}
\begin{subfigure}{0.2\textwidth}
\includegraphics[width=\linewidth]{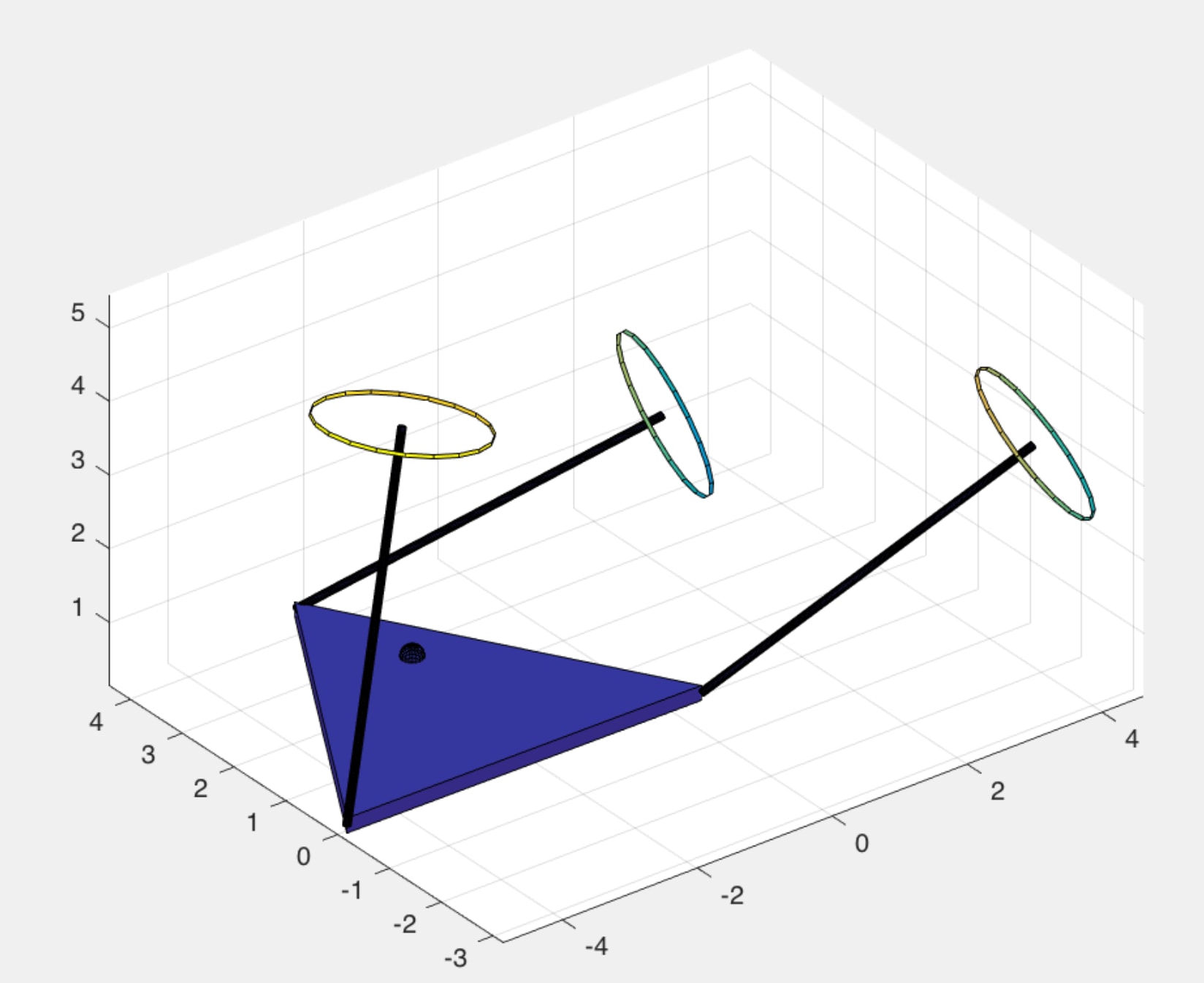}
\caption{}
\end{subfigure}

\medskip
\begin{subfigure}{0.2\textwidth}
\includegraphics[width=\linewidth]{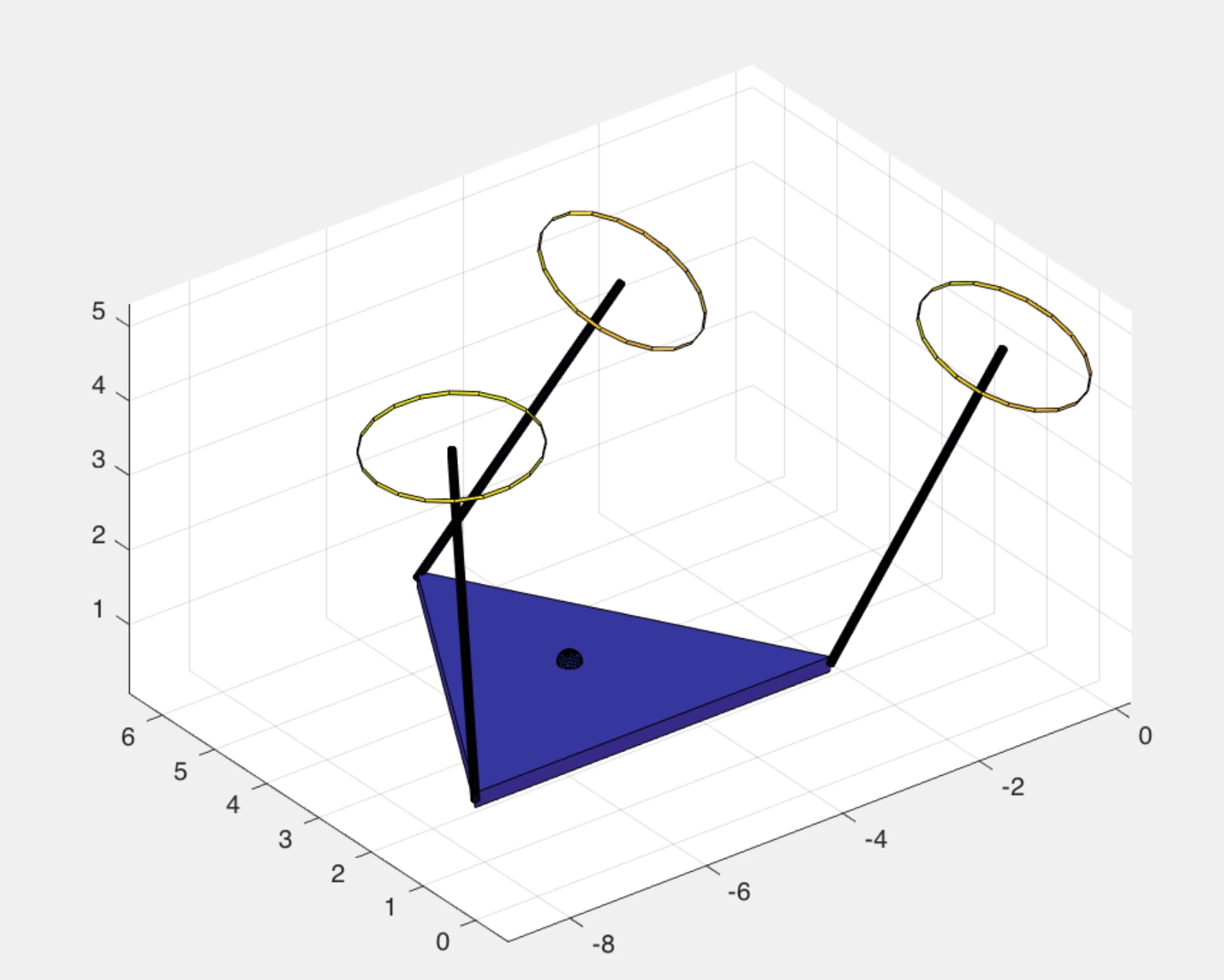}
\caption{}
\end{subfigure}
\begin{subfigure}{0.2\textwidth}
\includegraphics[width=\linewidth]{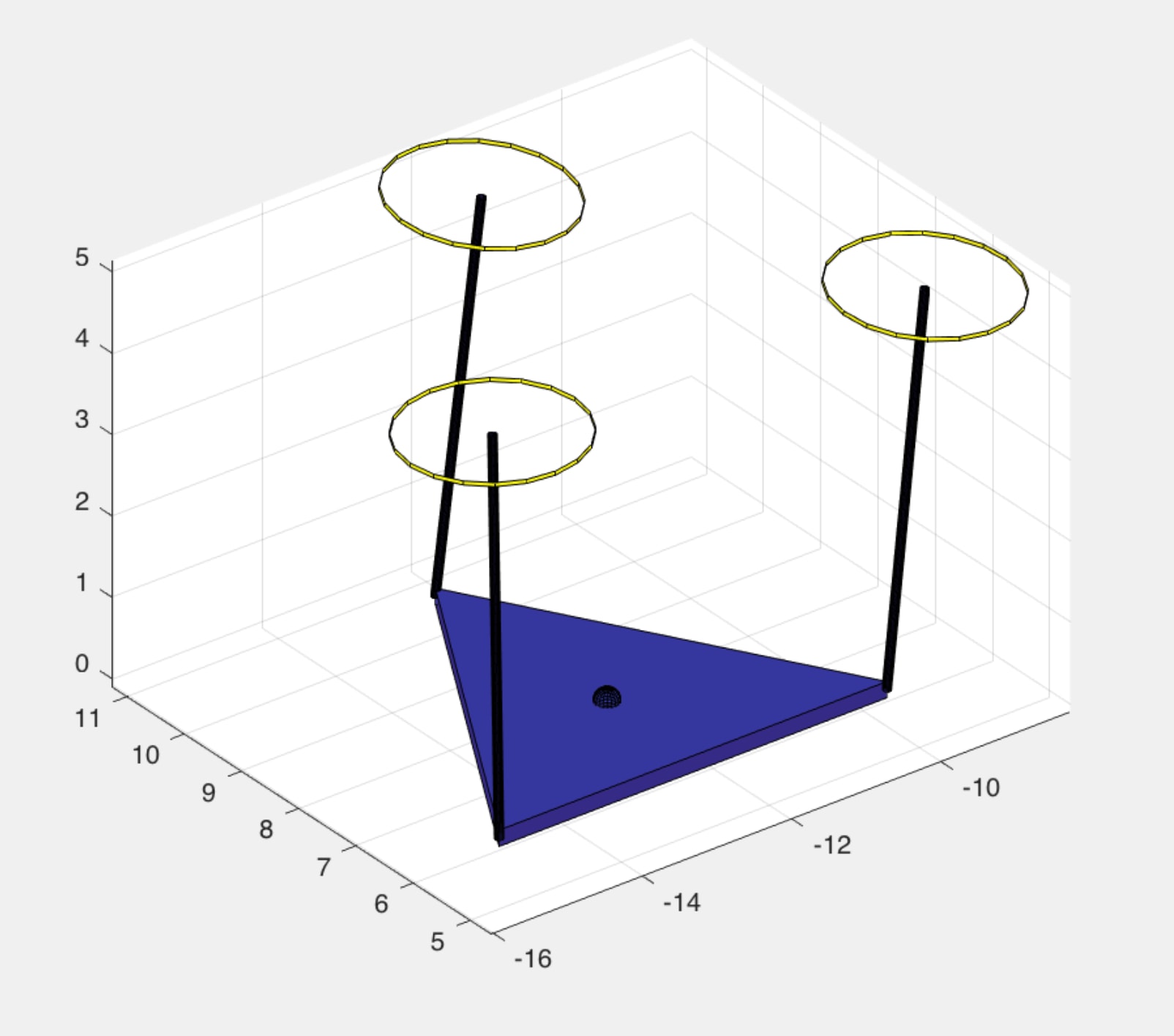}
\caption{}
\end{subfigure}

\caption{Pictorial depiction of the system with the proposed control strategy in effect. The animation is available at: \url{https://youtu.be/nGNS-eZxbVM}} \label{ani}
\end{figure}
The system trajectory subject to these simulation conditions, are depicted by the following plots
\begin{figure}[H]
\subcaptionbox{The plate's attitude quaternion}{\includegraphics[scale=0.11]{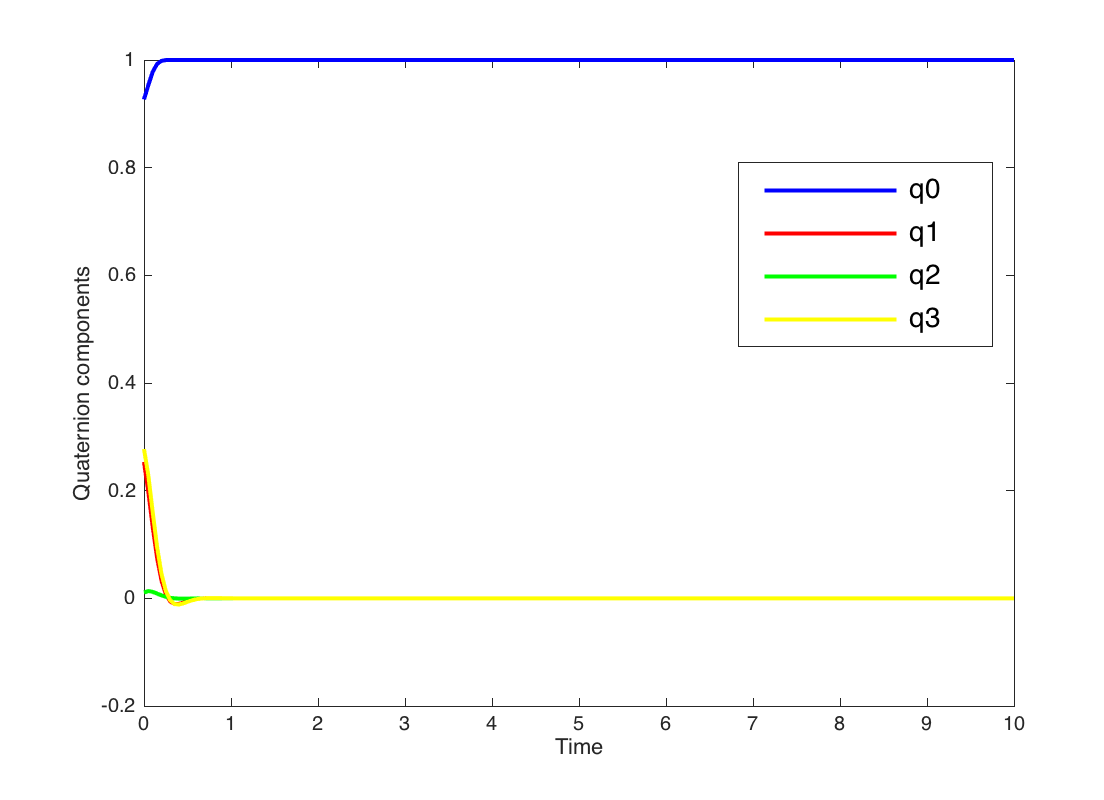}}%
\subcaptionbox{The plate's angular velocity}{\includegraphics[scale=0.11]{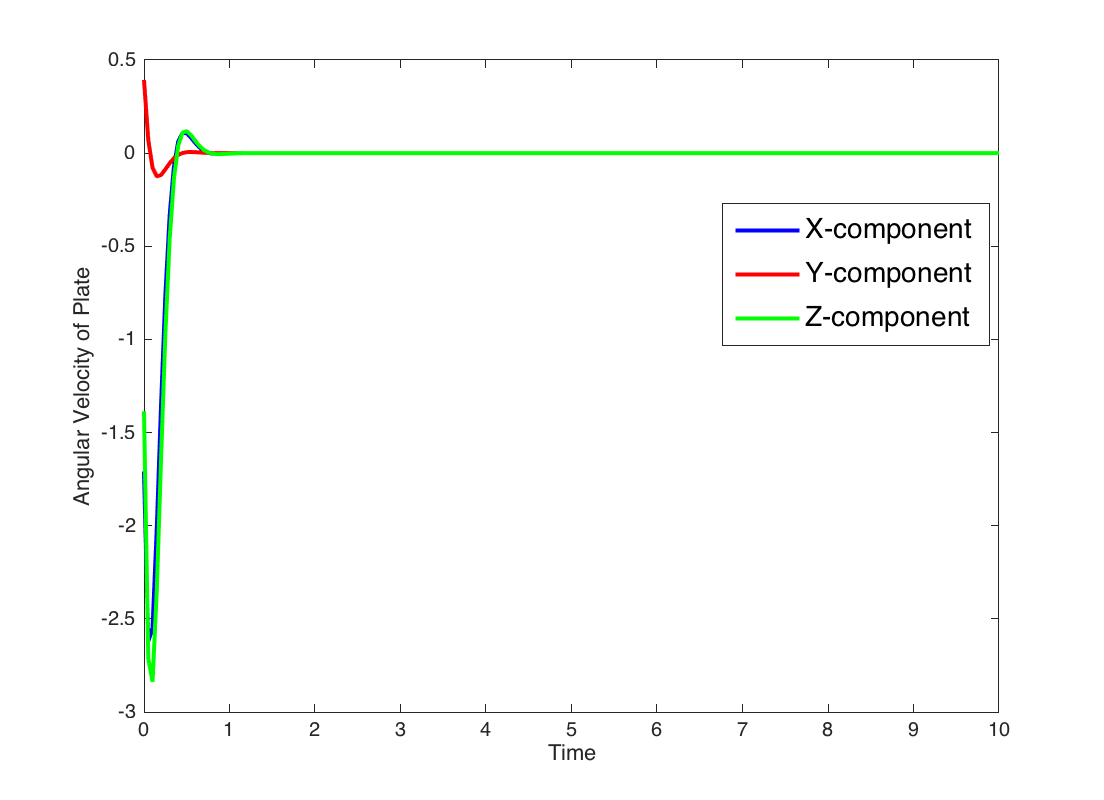}}
\caption{}
\end{figure}

\begin{figure}[H]
\subcaptionbox{The plate's position\label{pp}}{\includegraphics[scale=0.11]{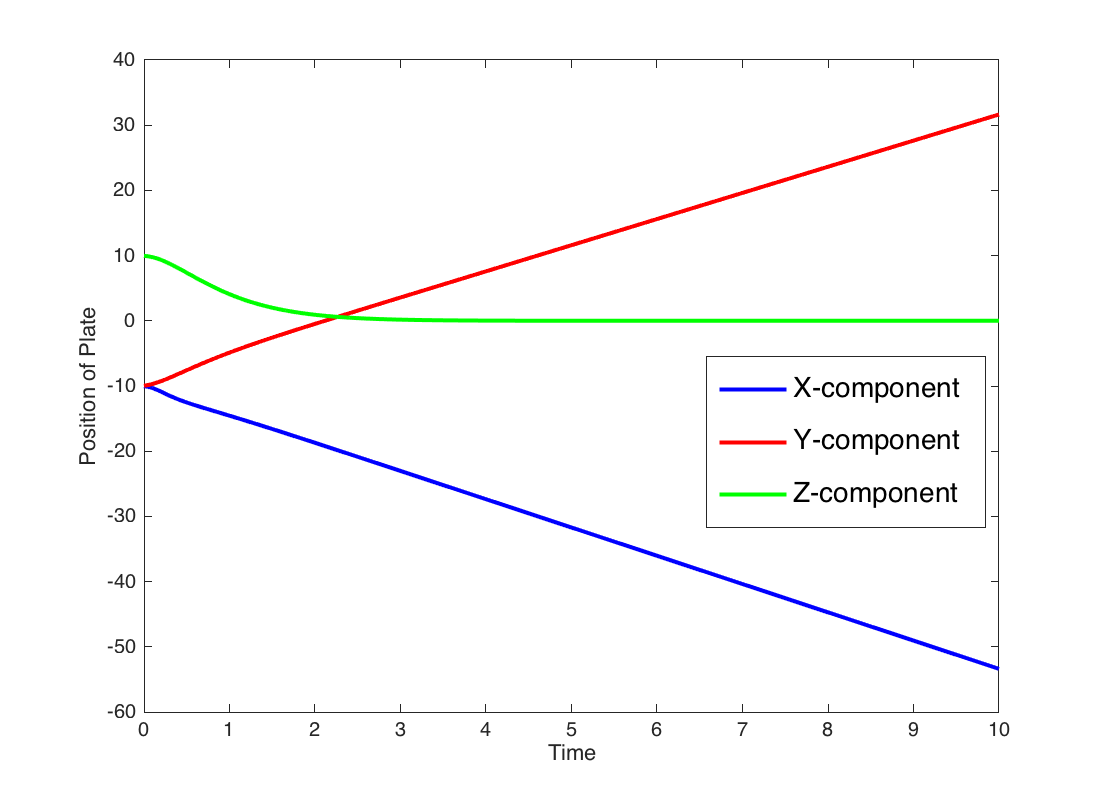}}%
\subcaptionbox{The plate's velocity\label{vp}}{\includegraphics[scale=0.11]{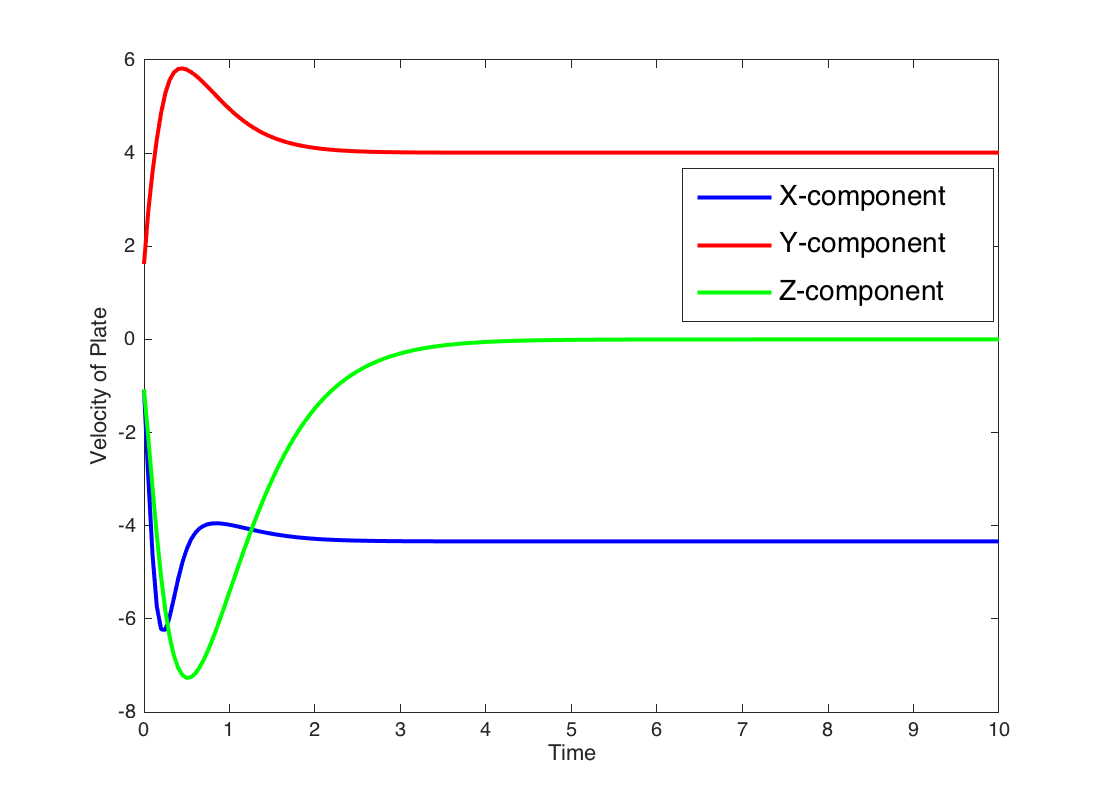}}\\
\caption{}
\end{figure}

Figures \ref{pp} and \ref{vp} show that the position of the plate in the x-y plane becomes unbounded while the velocity attains a constant value. This behaviour is attributed to the underactuated nature of the system.
\begin{figure}[H]
\subcaptionbox{The ball's position}{\includegraphics[scale=0.11]{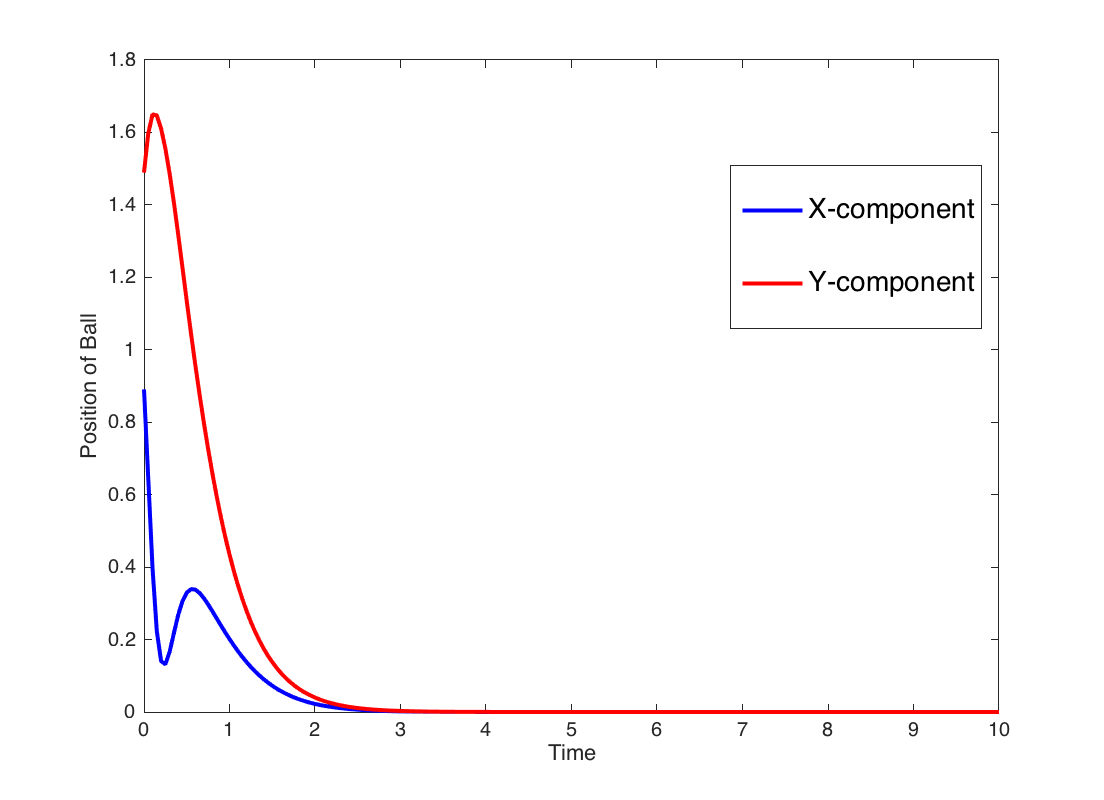}}%
\subcaptionbox{The ball's velocity}{\includegraphics[scale=0.11]{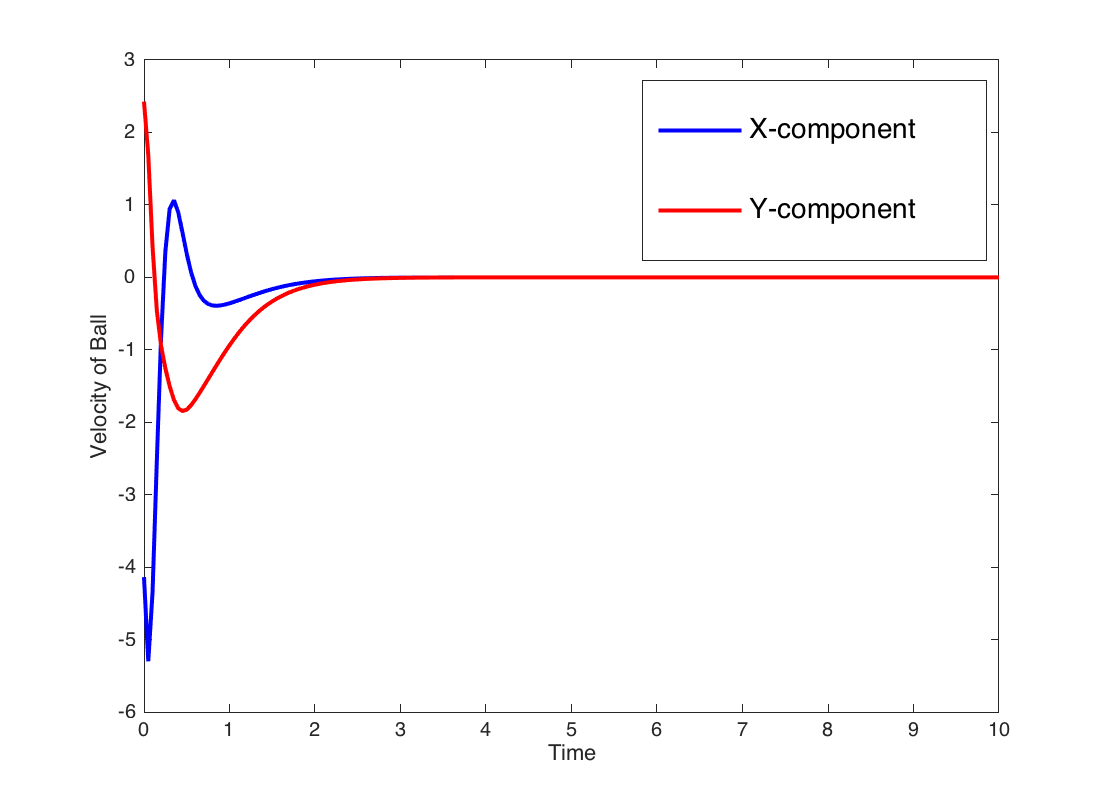}}\\
\caption{}
\end{figure}


 \bibliographystyle{IEEEtran}
\bibliography{IEEEabrv,IEEEexample}

\end{document}